\documentclass[aps,prd,twocolumn,showpacs,showkeys,superscriptaddress]{revtex4-1}

\usepackage{dcolumn}% Align table columns on decimal point
\usepackage{bm}% bold math
\usepackage{amssymb}
\usepackage{amsfonts}
\usepackage{amsmath}
\usepackage{graphicx}
\usepackage{epstopdf}
\usepackage{float}
\usepackage{relsize}
\usepackage{mathtools}
\usepackage{placeins}
\usepackage{soul}
\usepackage{ulem}
\usepackage{comment}
\usepackage{natbib}
\begin{document}

\title{Generalized Rastall's gravity and its effects on compact objects}% Force line breaks with 
\author{Cl\'esio E. Mota}
\email{clesio200915@hotmail.com}

\affiliation{Departamento de F\'isica, CFM - Universidade Federal de \\ Santa Catarina; C.P. 476, CEP 88.040-900, Florian\'opolis, SC, Brasil}

\author{Luis C. N. Santos}
\email{luis.santos@ufsc.br}

\affiliation{Departamento de F\'isica, Universidade Federal da \\ Para\'iba, C.P.5008, CEP 58.059-900, Jo\~ao Pessoa, PB, Brasil}

\author{Franciele M. da Silva}
\email{franmdasilva@gmail.com}

\affiliation{Departamento de Física, CCE - Universidade Federal \\ do Espírito Santo; CEP 29.075-910, Vit\'oria, ES, Brasil}

\author{Guilherme Grams}
\email{grams.guilherme@gmail.com}

\affiliation{Univ. Lyon, Univ. Claude Bernard Lyon 1, CNRS/IN2P3, IP2I Lyon, UMR 5822, F-69622, Villeurbanne, France}

\author{Iarley P. Lobo}
\email{iarley_lobo@fisica.ufpb.com}
\affiliation{Department of Chemistry and Physics, Federal University of Para\'iba, Rodovia BR 079 - Km 12, 58397-000 Areia-PB,  Brazil}

\affiliation{Departamento de F\'isica, Universidade Federal de \\ Lavras; C.P. 3037, CEP  37.200-900, Lavras, MG, Brasil.}

\author{D\'ebora P. Menezes}
\email{debora.p.m@ufsc.br}

\affiliation{Departamento de F\'isica, CFM - Universidade Federal de \\ Santa Catarina; C.P. 476, CEP 88.040-900, Florian\'opolis, SC, Brasil}

\begin{abstract} 
We present a generalization of Rastall's gravity in which the conservation law of the energy-momentum tensor is altered, and as a result, the trace of the energy-momentum tensor is taken into account together with the Ricci scalar in the expression for the covariant derivative. Afterwards, we obtain the field equations in this theory and solve them by considering a spherically symmetric space-time. We show that the external solution has two possible classes of solutions with spherical symmetry in the vacuum in generalized Rastall's gravity, and we analyse one of them explicitly. The generalization, in contrast to constant value $k=8\pi G$ in general relativity, has a gravitational parameter $k$ that depends on the Rastall constant $\alpha$. As an application, we perform a careful analysis of the effects of the theory on neutron stars using realistic equations of state (EoS) as input. Our results show that important differences on the profile of neutron stars are obtained within two representatives EoS.

\end{abstract}

\keywords{Generalization; Rastall's Gravity; Neutron Stars}

%\pacs{123}
\maketitle

\preprint{}

\volumeyear{} \volumenumber{} \issuenumber{} \eid{identifier} \startpage{1} %
\endpage{}
\section{Introduction}

Although general relativity (GR) has been successfully tested in many aspects, some open problems exist in both cosmology and astrophysics. Since the discovery of the discrepancy between the predicted rotation curves of galaxies and the observed motion \cite{vera}, and the ``missing mass'' of galaxy clusters \cite{cluster}, the dark matter hypothesis remains open. Moreover, the accelerated expansion of the universe observed today suggests the existence of the so-called dark energy \cite{Perlmutter:1998np,Sahni:2004ai}. Modified theories of gravity have gained attention because they may offer a way to solve these problems considering that these exotic forms of matter and energy are effects of a generalization of the GR due to a modified gravity. 

In this sense, the Rastall's theory of gravity, which may be obtained through a reinterpretation of the conservation  law  on  the energy-momentum tensor in curved spaces, couples the geometry to the matter in a modified way.
Rastall argued that the usual conservation law on the energy-momentum tensor $T_{\: \:\:\nu;\:\mu}^{\mu}=0$ is tested only in the Minkowski space-time such that in curved space-time it is possible to generalize this expression to $T_{\: \:\:\nu;\:\mu}^{\mu}=A_{\nu}$, where the functions $A_{\nu}$ vanish in flat space-time. Indeed, one possible implication of the modified conservation law in Rastall's gravity is to see the condition $T_{\: \:\:\nu;\:\mu}^{\mu}\neq 0$ as a consequence of the creation of particles in a cosmological context \cite{rastall1}. In astrophysics, the extra degree of freedom due to the modified expression for the divergence of the energy-momentum tensor has been explored in the study of neutron stars \cite{black2,santos8}, where the authors concluded that substantial modifications for the mass-radius relation are obtained even for very small alterations on the parameter of the Rastall's theory. We refer the reader to the following review on non-conservative theories of gravity \cite{Velten:2021xxw}.

Although the field equations in Rastall's theory are a generalization of the field equations in GR, it is well known that the static spherical symmetry 
solution in vacuum obtained with Rastall's theory coincides with the vacuum solution in GR \cite{black3}. In fact, it has been shown that there are two possible classes of solutions with spherical symmetry in vacuum in the Rastall's gravity. The first class of solutions is completely equivalent to the Schwarzschild solution while the second class of solutions has the same structure of the Schwarzschild--de Sitter solution in the GR \cite{black3}. But the effect of Rastall's theory is more evident and interesting in the presence of matter or electric charge, that is, $T_{\mu\nu}\neq 0$. 
Several works involving charged static spherically symmetric black holes and black hole solutions surrounded by fluid \cite{black4,black5,black6,black7,black8,black9,black10}, cosmological problems \cite{rastall2,rastall3,rastall4,rastall5,rastall6}, and other theoretical works \cite{rastall7,rastall8,rastall9,rastall10} have been explored in the Rastall's theory.

Some authors point out difficulties in describing the accelerated expansion of the universe from the usual Rastall gravity \cite{rastall2,Batista:2010nq}, which indicates the need for generalizing Rastall's approach \cite{Moradpour:2017shy,Lin:2020fue}. In this work, we intend to study a more general modification of the conservation law not yet explored in the usual Rastall gravitational theory. In his original work, Rastall has already mentioned the possibility of relating the conservation law of the energy-momentum tensor to $A_{\nu}$ %in the form  $T_{\: \:\:\nu;\:\mu}^{\mu}=A_{\nu}$ and then choose the following equation 
with
$A_{\nu}=\lambda \delta_{\:\:\nu}^{\mu}R_{,\:\mu}$. A generalization of this approach could be cast in the form of a general function of the Ricci scalar as $A_{\nu}=\delta_{\:\:\nu}^{\mu}f(R)_{,\:\mu}=f'(R)\delta_{\:\:\nu}^{\mu}R_{,\:\mu}$ (where $f'(R)\doteq df/dR$), which can be described by the usual Rastall equations, however endowed with a variable Rastall parameter. This approach has been explored in the literature (see, for instance \cite{Moradpour:2017shy}). However, inspired by $f(R,T)$ modifications of gravity \cite{frt,Houndjo:2011tu,Odintsov:2013iba}, it is possible to choose a function $A_{\nu}$ that depends on the scalar $R$ and, additionally, on the trace of the energy-momentum tensor $T$ in a more general way. Due to the fact that $R$ is associated to the modification of the conservation law, it is natural to assume that the coupling between $R$ and $T$ could also contribute to breaking the conservation law. We propose to choose a general function $A_{\nu}$ as the combination $A_{\nu}=(\alpha \delta_{\:\:\nu}^{\mu}R + \beta \delta_{\:\:\nu}^{\mu} RT)_{;\:\mu}$. Indeed, we explore this modification of the conservation law and its effects on the resulting solutions of the modified field equations. Similarly, some categories of modified theories of gravity as the $f(R,T)$ theory have considered field equations that depends on a function of $R$ and $T$ in which the trace could be induced by exotic imperfect fluids or quantum effects \cite{frt}.

In order to test the theory we use a well known astrophysical lab, the neutron stars. The death of a massive star in a core-collapse supernova can leave as remnant a neutron star or a black hole. A typical neutron star has about 1.4 $M_\odot$, a radius of the order of $11 - 13$ km and produces a strong gravitational field that can be used to test new gravity theories in extreme conditions. Additionally to the theoretical point of view, new experiments and observations like the NICER mission \cite{Miller_2019,Bogdanov_2019a,Bogdanov_2019b}, the LIGO-Virgo gravitational waves observations from neutron star merges \cite{PhysRevLett.121.161101,Abbott_2020} together with their electromagnetic counterpart \cite{Soares_Santos_2017,Valenti_2017} are making the astrophysical constraints to these objects continuously more restrictive, which makes these compact objects even more suitable to be used in tests of alternative gravity theories.

This paper is organized in the form: In Sec. \ref{1} we review the Rastall's theory of gravity and then expand the original work by considering a conservation relation that depends on the trace of energy-momentum tensor and on the Ricci scalar. We obtain the Newtonian limit of the field equations and study vacuum solutions with spherical symmetry. Neutron stars are considered in Sec. \ref{2} where we analyze the effects of the modified gravity on neutron stars mass and radius profiles using a soft and stiff realistic EoS. Finally, in Sec. \ref{3} we show our results.

\section{Generalization of Rastall's theory of gravity}
\label{1}

In order to expand the original work and consider a conservation relation that depends on the  trace of energy-momentum tensor, we will briefly review the original theory \cite{Rastall}. Then, we will show how to modify Einstein field equations such that the non-conservative aspect of generalized theory will be taken into account.

\subsection{Rastall's theory} 

The left hand side of the usual Einstein field equations satisfies $G_{\: \:\:\nu;\:\mu}^{\mu}=0$, which may be easily verified by using the Bianchi identities. In fact, this relation is in accordance with the right hand side of the field equations if one considers the conservation law $T_{\: \:\:\nu;\:\mu}^{\mu}=0$. However, in Rastall's gravity \cite{Rastall} it is argued that this equation, in a general space-time, may be replaced by the modified relation $T_{\: \:\:\nu;\:\mu}^{\mu}=\lambda R_{,\:\nu}$ where $\lambda$ is a constant. After rewriting the terms of this equation we obtain the relation \begin{equation}
\left(T_{\: \: \mu}^{\nu}-\lambda\delta_{\: \:\mu}^{\nu}R\right)_{;\: \nu} =0.
\label{eq1}
\end{equation}  
In this way, Eq. (\ref{eq1}) can be used to generalize the Einstein field equations so that the term in brackets in this equation is used on the right side of the field equations. The result is
\begin{equation}
R^{\nu}_{\: \: \mu}-\frac{1}{2}\delta_{\: \:\mu}^{\nu}R= k\left(T_{\: \: \mu}^{\nu}-\lambda\delta_{\: \:\mu}^{\nu}R\right), \label{eq02}
\end{equation}
where $k$ is the Rastall coupling constant. 

\subsection{Generalized Rastall's theory}

We discuss now the generalization of Rastall's theory of gravity. We propose that the general function 
$A_{\nu}$ related to the divergence 
of the energy-momentum tensor in 
curved space-time is given by
$A_{\nu}=(\alpha \delta_{\:\:\nu}^{\mu}R + \beta \delta_{\:\:\nu}^{\mu} RT)_{;\:\mu}$,
{\it i.e.}, it has the same dependence on $R$ as in the original Rastall work, in addition to a coupling term defined by $RT$, where $T$ is the trace  of  the  energy-momentum tensor $T_{\mu\nu}$. In particular, it is expected that the final form of the field equations in this theory will incorporate the elements of this modification and will be able to reproduce the main features of Rastall's gravity in a particular case. 
As a test for the theory, we solve the field equations that originates from a metric that can be used to model space-time compact stars, such as neutron stars, and thus analyze the possible effects on the mass versus radius diagrams of these objects.
As mentioned, the modification in the energy-momentum conservation law has the following form:
\begin{equation}
T_{\: \:\:\nu;\:\mu}^{\mu}=\alpha R_{,\:\nu}+\beta\left(RT\right)_{,\:\nu},
\label{eq2}
\end{equation}
where $\alpha$ and $\beta$ are called coupling parameters, which measure the deviation from standard theory of GR and quantify the affinity of the matter field coupled with geometry. 

The usual Rastall's gravity can be recovered in the appropriate limit of $\beta \rightarrow 0$. The divergence of $T_{\mu\nu}$ given by equation (\ref{eq2}) is proportional to the gradients of $R$ in both terms. Therefore, in the flat space-time, when $R=0$, the usual conservation law is recovered. From equation (\ref{eq2}) we implement the following expression:
\begin{equation}
\left(T_{\: \: \mu}^{\nu}-\alpha\delta_{\: \:\mu}^{\nu}R-\beta\delta_{\: \:\mu}^{\nu}RT\right)_{;\: \nu} =0. \label{eq4}
\end{equation} 

In fact, assuming the condition given by the above expression, the modified field equations of generalized Rastall's gravity can be written as
\begin{equation}
R^{\nu}_{\: \: \mu}-\frac{1}{2}\delta_{\: \:\mu}^{\nu}R= k\left(T_{\: \: \mu}^{\nu}-\alpha\delta_{\: \:\mu}^{\nu}R-\beta\delta_{\: \:\mu}^{\nu}RT\right),
\label{eq5}
\end{equation}
where $k$ is the modified gravitational coupling constant in this theory. Taking the trace of the previous equation, we have 
\begin{equation}
R=\frac{kT}{4k\left(\alpha+\beta T\right)-1}, \label{eq6}
\end{equation}
which leads to the following expression
\begin{equation}
R_{\mu\nu}-\frac{1}{2}g_{\mu\nu}R=k\tau_{\mu\nu} \label{eq7},
\end{equation}
where $\tau_{\mu\nu}$ is called effective energy-momentum tensor having the following expression \footnote{Since we assume small departures from GR, divergences arising from this description in terms of an effective energy-momentum tensor shall not emerge at the scale of neutron stars analyzed in this paper.}:
\begin{equation}
\tau_{\mu\nu}=T_{\mu\nu}-\frac{g_{\mu\nu}T}{4-\frac{1}{k\left(\alpha+\beta T\right)}}. \label{eq8}
\end{equation}
In the next section, we use the Newtonian limit to obtain in this context the form of the gravitational constant $k$.

Notice that, $T_{\mu\nu}$ is the energy-momentum tensor of ordinary matter (built from the matter fields). 
\par
Although we are able to write the Generalized Rastall's field equations as the Einstein tensor sourced by a function of the energy-momentum tensor (an effective ``energy-momentum tensor''), this approach presents non-trivial differences in comparison to Einstein's theory as shall be seen in the next sections.
\par
A possible equivalence between the original Rastall's approach and Einstein's theory has been originally claimed in \cite{Lindblom} (and more recently in \cite{Visser:2017gpz}) based on the rearrangement of the field equations leading to the conservation of an effective energy-momentum tensor, a property that could still persist in our generalized proposal. However, it has been shown in \cite{Darabi:2017coc} that the assumption that Einstein and Rastall's theories present the same energy-momentum tensor, i.e, built from the matter fields, implies in the physical unequivalence of these theories (the same argument is valid in our case). In fact, the Lagrangian whose variation gives rise to Rastall's gravity has also been a matter of current research. Due to the covariant non-conservation of the stress-energy tensor, it is expected that some kind of non-minimal coupling between matter and geometry could generate this kind of theory. In fact, some comments about this issue can be found in \cite{Darabi:2017coc,Lobo:2020jfl} and the subject has been  analyzed in \cite{Smalley,DeMoraes:2019mef,Santos:2017nxm}, but a final word has not been stated on the matter (see review \cite{Velten:2021xxw}). Hence, we do not have a Lagrangian for the present generalized version of Rastall gravity either. We expect a non-trivial coupling between matter and the Ricci tensor could be a candidate for generating the new terms in this new proposal.

\subsection{Newtonian limit}

Next we calculate the Newtonian limit of Einstein field equations so that we can obtain the value of the constant $k$ in our generalized Rastall's theory. To do this, we compare our field equations in the weak field regime with the Poisson's equation:
\begin{equation}
\nabla^{2}\phi=4\pi G\rho .\label{eq9}
\end{equation}
In the Newtonian limit we can replace the metric tensor $g_{\mu\nu}$ by the Minkowski tensor $\eta_{\mu\nu}$, in terms that multiply the curvature, so that equation (\ref{eq5}) reads:
\begin{equation}
R_{\mu\nu}-\frac{1}{2}\eta_{\mu\nu}R=kT_{\mu\nu}-k\eta_{\mu\nu}\left(\alpha R+\beta RT\right). \label{eq10}
\end{equation}
In this limit we have $\rho\gg p$ and therefore $\left|T_{00}\right|\gg\left|T_{ij}\right|$ \cite{weinberg}, so that if we look at the $\left(00\right)$ component of equation (\ref{eq10}) we find the following:
\begin{equation}
R_{00}-\frac{1}{2}(-1)R=k(-\rho)-k(-1)\left(\alpha R+\beta R(-\rho)\right). \label{eq11}
\end{equation}
Using the approximation $R\approx \sum_{k=1}^3 R_{kk}-R_{00}$, we can obtain the relation:
\begin{equation}
R=\frac{2R_{00}}{1-6k\alpha+6k\beta\rho}
\end{equation}
using this relation in equation (\ref{eq11}) and knowing that $R_{00}\approx-\nabla^{2}\phi$, we will find:
\begin{equation}
k\left(\frac{1-6k\alpha+6k\beta\rho}{1-4k\alpha+4k\beta\rho}\right)=8\pi G
\end{equation}
When solving this equation for $k$, we obtain two possible solutions. However, one of these solutions diverges when $\alpha \rightarrow 0$, so that the physically acceptable solution is given by the following expression:
\begin{align}
k=&\frac{1+32\pi G\left(\alpha-\beta\rho\right)}{12\left(\alpha-\beta\rho\right)} \nonumber \\
-&\frac{\sqrt{1+32\pi G\left(\alpha-\beta\rho\right)\left(32\pi G\left(\alpha-\beta\rho\right)-4\right)}}{12\left(\alpha-\beta\rho\right)}\, .
\label{kRgen}
\end{align}

In principle, this equation would indicate the presence of a $\rho$-dependent gravitational coupling constant, $k$, induced by $\beta$-dependent contributions. We could avoid this feature by an analysis of $k$ in some limiting situations by restricting the range of $\beta$ such that even for the extreme regime in which Newtonian gravity should be valid, like considering the maximum density of white dwarfs $\rho^{\text{max}}_{\text{wd}}\approx 10^{7}\, g/cm^3=2.8\times 10^{-8}fm^{-4}$ (in natural units $\hbar=c=1$), the dimensionless condition $128\pi G \rho^{\text{max}}_{\text{wd}} |\beta|\ll 1$ is valid. This sets a condition that suppresses pure-$\beta$ contributions in the determination of $k$ from the Newtonian limit even when $\alpha$-terms are absent. This requirement can be derived by expanding the expression in the square root of (\ref{kRgen}) as
\begin{widetext}
\begin{equation}\label{square}
1+128 \pi  G \beta \rho +1024 \pi ^2 G^2 \beta ^2 \rho ^2-2048 \pi ^2 G^2 \alpha  \beta   \rho+ 1024 \pi ^2 G^2 \alpha ^2 -128 \pi  \alpha  G\, .    
\end{equation} 
\end{widetext}
From this expression, one sees that even when $\alpha \rightarrow 0$, the terms proportional to $G\rho \beta$ are suppressed. In fact, in this limit we would have
\begin{widetext}
\begin{equation}
    k\xrightarrow{\alpha\rightarrow 0}\frac{8\pi G}{3}-\frac{1}{12\beta \rho}+\frac{\sqrt{1+128\pi G\beta\rho+1024\pi^2G^2\beta^2\rho^2}}{12\beta\rho}\xrightarrow{G|\beta|\rho\ll 1} 8\pi G\, .
\end{equation}
\end{widetext}
Besides that, if $G\alpha$ and $G\beta\rho^{\text{max}}_{\text{wd}}$ are at most of the same order of magnitude, then $k\approx 8\pi G$, since $G\beta\rho\ll 1$ in the Newtonian regime.
\par
But is there an upper limit to the magnitude of $G \alpha$ terms? Now, we check again the term in the square root of (\ref{kRgen}), given by (\ref{square}), to analyze the general $\alpha$ contribution and its relation with $\beta$, where in this case, we see the presence of a term that couples these parameters, given by $2048 \pi ^2 G^2 \alpha  \beta   \rho$ that could eventually lead an amplification of this term and the presence of a density-dependence of the coupling constant (\ref{kRgen}). So, we should expect that a bound on $|\alpha|$ would need to be set in order to avoid an amplification of this contribution. Since $2048 \pi ^2 G^2 \alpha  \beta   \rho=16\pi G\alpha(128\pi G \beta\rho)$, we should have $G|\alpha|\not\gg 1$: for instance, a sufficient condition would be $G|\alpha|\leq 1$, but obviously this is not a necessary condition, one should only be aware that the dimensionless quantity $2048 \pi ^2 G^2 |\alpha|  |\beta|   \rho^{\text{max}}_{\text{wd}}$ does not approach $1$.
\par
In summary, if we assume $G|\beta|\rho^{\text{max}}_{\text{wd}}\ll 1$ and, for instance, $G|\alpha|\leq 1$ (where $\rho^{\text{max}}_{\text{wd}}\sim 10^{-8}fm^{-4}$) as conditions in our parameter space $(\alpha,\beta)$, then our Newtonian limit does not present a density-dependent coupling constant, it depends only on $\alpha$ and $G$ and is given by
\begin{equation}
k=\frac{1+32\alpha G\pi-\sqrt{1+32\alpha G\pi (32\alpha G\pi -4)}}{12\alpha}\, .   
\label{kRgenb0}
\end{equation}

Since $\rho^{\text{max}}_{\text{wd}}\approx 2.8\times 10^{-8}fm^{-4}$, then in a system of units in which $c=\hbar=1$, 
which implies that $G=6.67\times 10^{-11}\, m^3/kg\cdot s^2=2.6\times 10^{-40}\, fm^2$, we would have found $\beta\ll 10^{44}\, fm^2$ and $\alpha\leq 3.8\times 10^{-41}\, fm^{-2}$ as the conditions for a constant $k$. On the other hand, in a system of natural units in which $c=\hbar=G=1$, the above conditions are translated as $\beta\ll 10^5\, fm^4$ and $\alpha\leq 1$. This last system of units will be used later on in the last section of this paper, when we  consider a region of the parameter space $(\alpha,\beta)$ %according to these restrictions,
such that 
%we can use 
the coupling constants are defined in (\ref{kRgenb0}), as derived from the Newtonian limit of the theory.

In addition, we can verify that taking the limit $\alpha\rightarrow0$, we  will recover, as would be expected, the value of $k$ for the GR, that is $k=8\pi G$ . As will be seen, we shall verify the impact of generalized Rastall parameters that obey these restrictions on the equation of state of neutron stars, which nevertheless shall leave perceivable imprints. In the next section, we will see that for solutions associated with the vacuum, there are two possible space-times with spherical symmetry.

\subsection{Vacuum solution with spherical symmetry}
\label{modTOV}

At this point, we are interested in solutions of the field equations that represent static spherically symmetric space-times in generalized Rastall's gravity. In the first place, we consider the trace of the field equations (\ref{eq5}) in vacuum: $R(-1+4k\alpha)=0$, this equation is satisfied either by setting $R=0$ or $\alpha = 1/4k$. In the first case, it is possible to show that the spherical symmetric solution is completely equivalent to the usual Schwarzschild solution in GR. In the second case, i.e., for $\alpha =1/4k$, equation (\ref{eq6}) indicates that the Ricci scalar is independent of matter distribution what could be inconsistent with EoSs related to compact stars. Hence, we next consider only the first solution. We refer the interested reader to a detailed analysis of the second type of solution in Rastall's gravity, which  can be found in \cite{black3}.

We observe that the vacuum version of Eq. (\ref{eq5}) reads
\begin{equation}
R_{\: \:\mu}^{\nu}-\frac{1}{4}\delta_{\: \:\mu}^{\nu}R=0,
\label{eqc1}
\end{equation}
and the metric on the symmetry of interest, can be written in the usual form 
\begin{equation}
ds^{2}=-B(r)dt^{2}+A(r)dr^{2}+r^{2}(d\theta^{2}+\sin{\theta}^{2}d\phi^{2}),
\label{ds2}
\end{equation}
where $B(r)$ and  $A(r)$ are functions that are determined using the field equations. In this way, if one uses the metric (\ref{ds2}) in the field equations (\ref{eqc1}), one obtain the following vacuum equations
\begin{align}
R_{tt}-g_{tt}\frac{1}{4}R=0, \label{eqc2}\\
R_{rr}-g_{rr}\frac{1}{4}R=0, \label{eqc3}\\
R_{\theta\theta}-g_{\theta\theta}\frac{1}{4}R=0, 
\label{eqc4} 
\end{align}
where 
\begin{align}
R_{tt} &= -\frac{1}{4}\left(\frac{B'A'}{A^2}+\frac{B'^2}{AB}\right)+\frac{1}{2}\frac{B''}{A} + \frac{B'}{rA},  \label{eqc5}\\
R_{rr} &= \frac{1}{4}\left(\frac{B'A'}{BA}+\frac{B'^2}{B^2}\right)-\frac{1}{2}\frac{B''}{B} + \frac{A'}{rA},  \label{eqc6}\\
R_{\theta\theta}&= \frac{1}{2}\frac{rA'}{A^2}-\frac{1}{2}\frac{B'r}{BA}+1-\frac{1}{A}. \label{eqc7}
\end{align}
Solving the differential equations obtained from eq. (\ref{eqc1}) for $R=0$, we obtain the usual space-time with spherical symmetry in generalized Rastall's gravity which is equivalent to the Schwarzschild solution in GR. As we will see, the  expected effect of generalized Rastall's gravity will be observed in the case of $T_{\mu\nu} \neq 0$.

\section{Neutron stars} \label{2}

The study of neutron stars is interesting from a nuclear physics point of view, thanks to the extremely dense matter and possible phase transitions inside the stars, as well as a good test for alternative theories of gravity, due to the intense gravitational field created by this object. 

Therefore, in order to test the generalized Rastall's gravity, in the next section we derive
the equations that describe neutron stars within this theory.

\subsection{Internal solution}

Here we present the solution of the modified Einstein equations for the interior of a compact, static and spherically symmetric object.

The distribution of matter inside the star can be described by the energy-momentum tensor of a perfect fluid, given by the following expression:

\begin{equation}
T_{\mu\nu}=pg_{\mu\nu}+(p+\rho)U_{\mu}U_{\nu}
\label{Tmunu},
\end{equation}
where $p$ and $\rho$ are respectively the pressure and the energy density of the stellar matter, and $U_{\mu}$ is the 4-velocity of the fluid element, which satisfies $U_{\mu}U^{\mu}=-1$.

Now we can work on Eq. (\ref{eq7}) together with the metric, Eq. (\ref{ds2}), and Eq. (\ref{Tmunu}) to obtain the components of the modified Einstein field equations  

\begin{equation}
-\frac{B}{r^{2}A} + \frac{B}{r^{2}} + \frac{A'B}{rA^{2}} = 8\pi GB \Bar{\rho},  \label{eq14}
\end{equation}

\begin{equation}
-\frac{A}{r^{2}} + \frac{B'}{rB} + \frac{1}{r^2}  = 8\pi GA \Bar{p},  \label{eq15}
\end{equation}

{\fontsize{9.7}{12}
	\begin{align}
	-\frac{B'^{2}r^{2}}{4AB^{2}} - \frac{A'B'r^{2}}{4A^{2}B} + \frac{B'' r^{2}}{2AB} - &\frac{A'r}{2A^{2}} + \frac{B'r}{2AB} \nonumber \\
	& = 8\pi Gr^{2} \Bar{p},  
	\label{eq16}
	\end{align}}

where the effective energy and pressure read

\begin{align}
\Bar{\rho} & = \frac{k}{8\pi G}\left[\rho+\frac{T}{4-\frac{1}{k\left(\alpha+\beta T\right)}}\right],\\
\Bar{p} & = \frac{k}{8\pi G}\left[p-\frac{T}{4-\frac{1}{k\left(\alpha+\beta T\right)}}\right],
\end{align}
with $T = 3p - \rho$. Note that $k$ in the above equations is given by Eq. (\ref{kRgenb0}), where we can see the effect of generalized Rastall's gravity.

From Eq. (\ref{eq14}) we can integrate $A$,

\begin{equation}
A(r)=\left[1-\frac{2GM(r)}{r} \right]^{-1}, \label{eq23}
\end{equation}
and $M(r)$ is the mass included in the radial coordinate $r$. The definition of the mass term is
\begin{equation}
M(r)=\int_{0}^{R} 4\pi r'^{2}\Bar{\rho}(r')dr',
\label{mass}
\end{equation}
where $R$ is the radius of the star, which is defined as the radial coordinate at which the pressure vanishes, {\it i.e.}, $R\equiv r'(p=0)$. 
Therefore, the total gravitational mass of the neutron stars is $M \equiv M(R)$.

We want to analyze the mass and radius of neutron stars using the pressure and energy density of the nuclear matter inside the star as inputs. The mass equation (\ref{mass}) is one of our equations, and the second one we obtain from a combination of Eqs. (\ref{eq15}) and (\ref{eq23}) to complete our system:

\begin{align}
\frac{B'}{2B} & = \frac{G M(r)}{r^{2}} \left[1+\frac{4\pi r^{3} \Bar{p}}{M(r)} \right]\left[1-\frac{2G M(r)}{r}\right]^{-1}. \label{eq22} 
\end{align}

The generalized Rastall's gravity directly affects the energy-momentum conservation, as explained in the previous section. 
Therefore, from the non-conservation of $T_{\: \: \mu}^{\nu}$ given by Eq. (\ref{eq2}), we obtain

\begin{equation}
\frac{B'}{2B}  = -\frac{\Bar{p}'}{\Bar{p}+\Bar{\rho}}. 
\end{equation}

We manipulate the last two equations to obtain the following relation:

\begin{equation}
\Bar{p}'= -\frac{G M(r)\Bar{\rho}}{r^{2}}\left[1+\frac{\Bar{p}}{\Bar{\rho}}\right] \left[1+\frac{4\pi r^{3} \Bar{p}}{M(r)} \right]\left[1-\frac{2G M(r)}{r}\right]^{-1}. 
\label{plinha}
\end{equation}

Equations (\ref{mass}) and (\ref{plinha}) are
the equivalent of the Tolman-Oppenheimer-Volkoff (TOV) \cite{tovtolman,tovvolk} equations in the generalized Rastall's gravity. In the next section we use these two equations together with the nuclear equation of state to obtain the mass and radius of a family of neutron star in the context of the modified theory of gravity presented in this work.

\subsection{Numerical results}\label{subsec:numerical}

We are now in a position that allows us to use the equations obtained in the previous section together with realistic equations of state to model neutron stars. All of the analyses done in this section are in units $c=\hbar=G=1$, so the parameter $\alpha$ is dimensionless and $\beta$ has units of $fm^4$.

As input to the stellar equilibrium equations, we use two realistic equations of state (EoS) obtained from a relativistic mean field (RMF) approach. We first consider
the IU-FSU \cite{PhysRevC.82.055803} parametrization 
because it is able to explain reasonably well both nuclear \cite{PhysRevC.99.045202} and stellar matter properties \cite{PhysRevC.93.025806}. We then compare the IU-FSU results with the ones obtained with a stiffer EoS calculated with the TM1 parametrization \cite{tm1}. It is well known that a stiffer EoS leads to a bigger NS maximum mass in contrast to a softer one. For the neutron star crust, we use the well known BPS \cite{BPS} EoS that describes well the low density region. 

\begin{figure}[h!]
	\centering
	\begin{tabular}{ll}
		\includegraphics[width=5.8cm,angle=270]{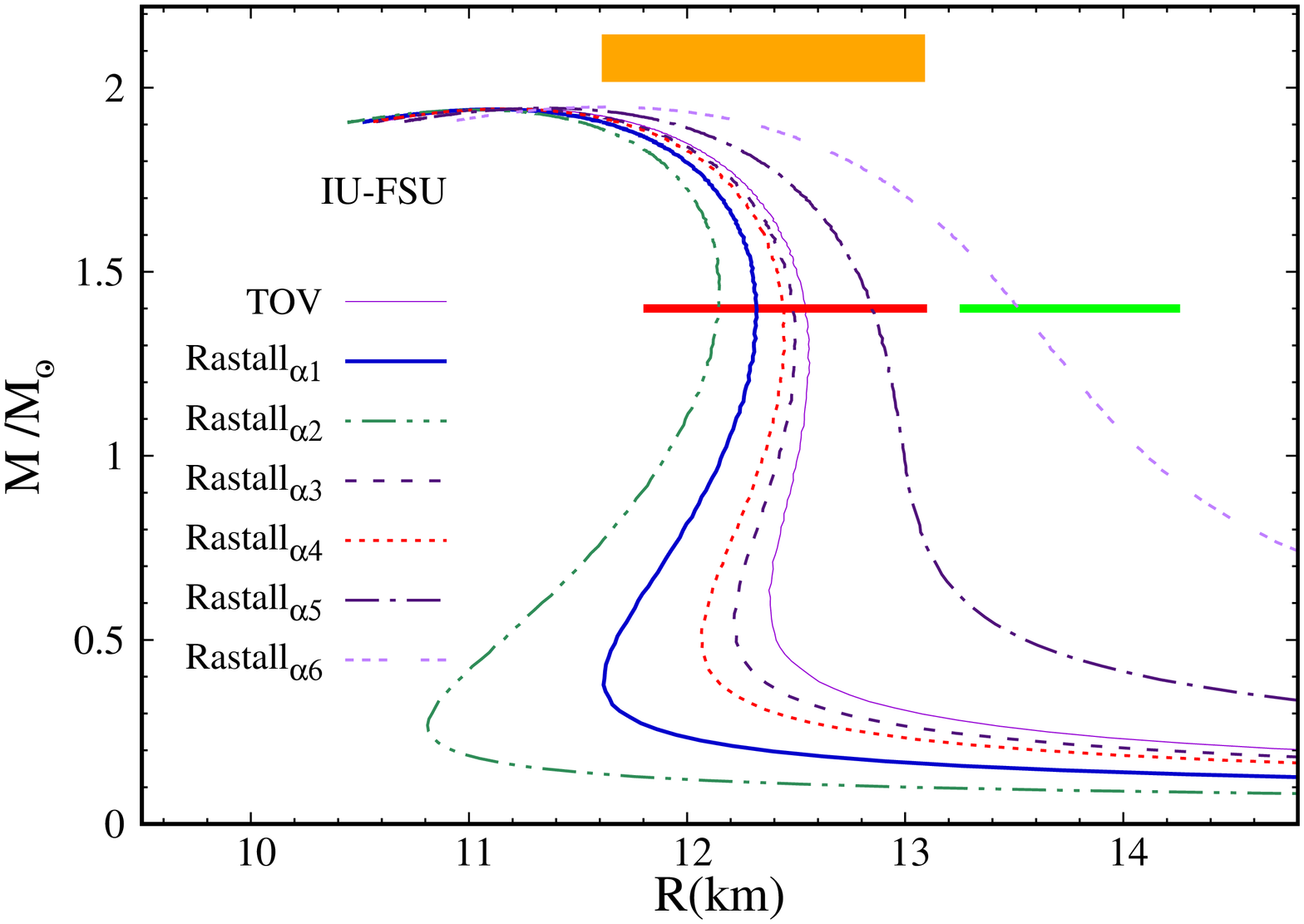}\\
		\includegraphics[width=5.8cm,angle=270]{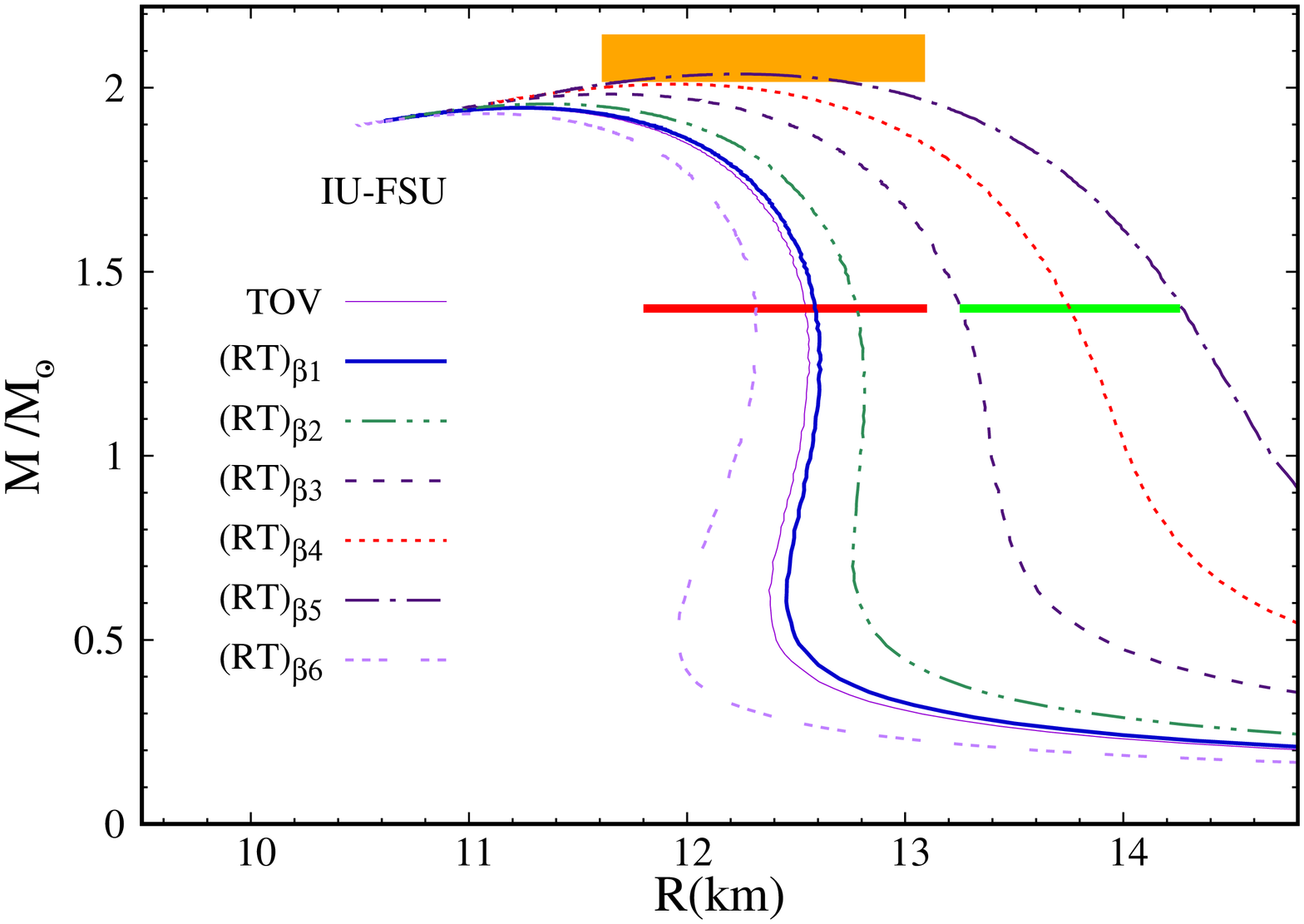} 
	\end{tabular}
	\caption{Mass-radius relation for a family of hadronic stars described with the IU-FSU EoS. We analyze the effects of varying the parameter $\alpha$ (top) while keeping the parameter $\beta$ null, i.e., here we have the original Rastall theory, and the effects of varying the parameter $\beta$ (bottom) while keeping $\alpha$ null, therefore in this case we study the $RT$ term alone. The red and green line segments correspond respectively to the radius range of the $1.4 M_\odot$ NS for PSR J0030 + 0451 and PREX-2. The orange rectangular region represents the interval of radius estimate for the $2.08 \pm 0.07 M_\odot$ NS PSR J0740+6620. See text for details.}
	\label{fig_IUFSU1}
\end{figure}

\begin{figure}[h!]
	\centering
	\begin{tabular}{ll}
		\includegraphics[width=5.8cm,angle=270]{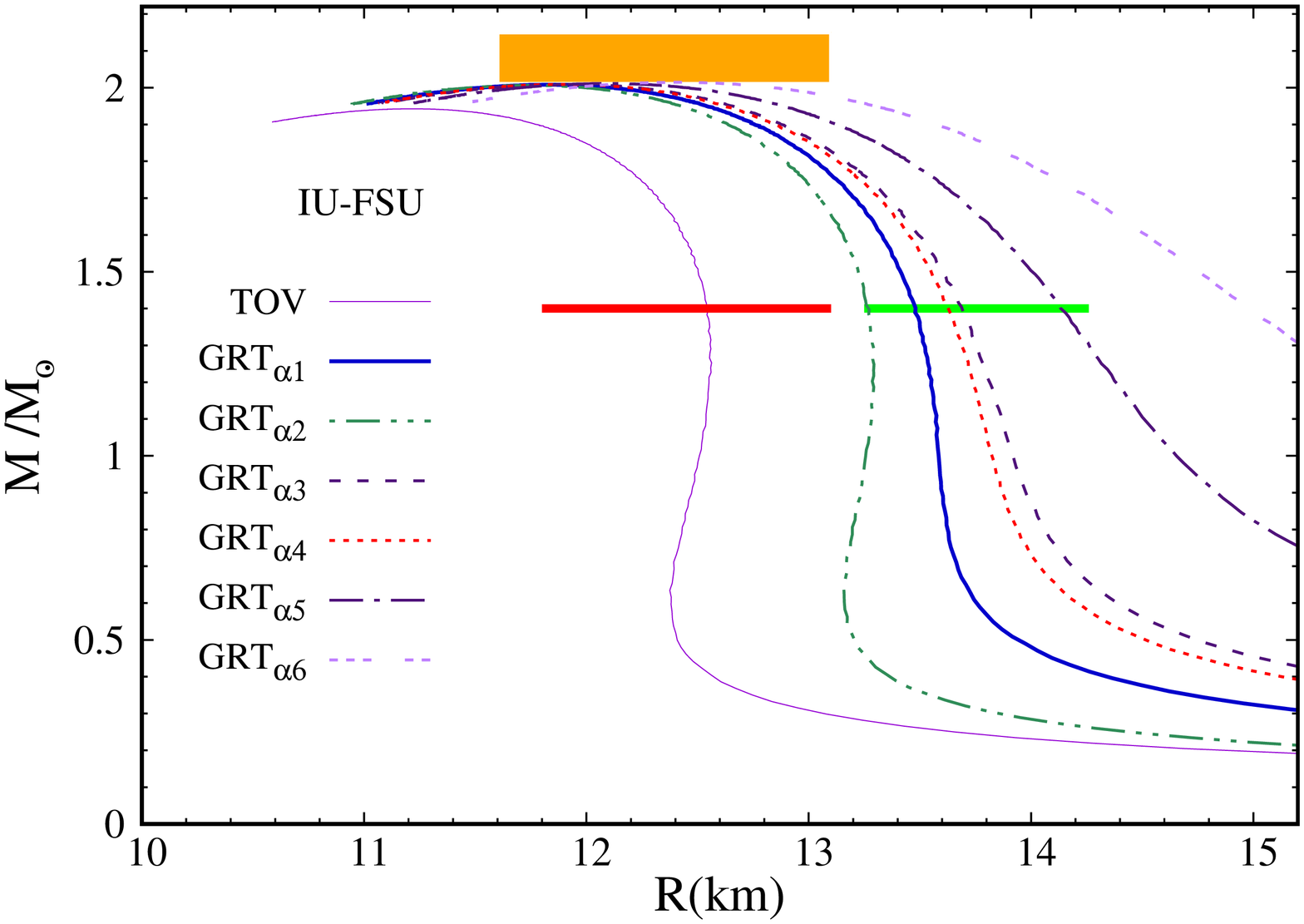}\\
		\includegraphics[width=5.8cm,angle=270]{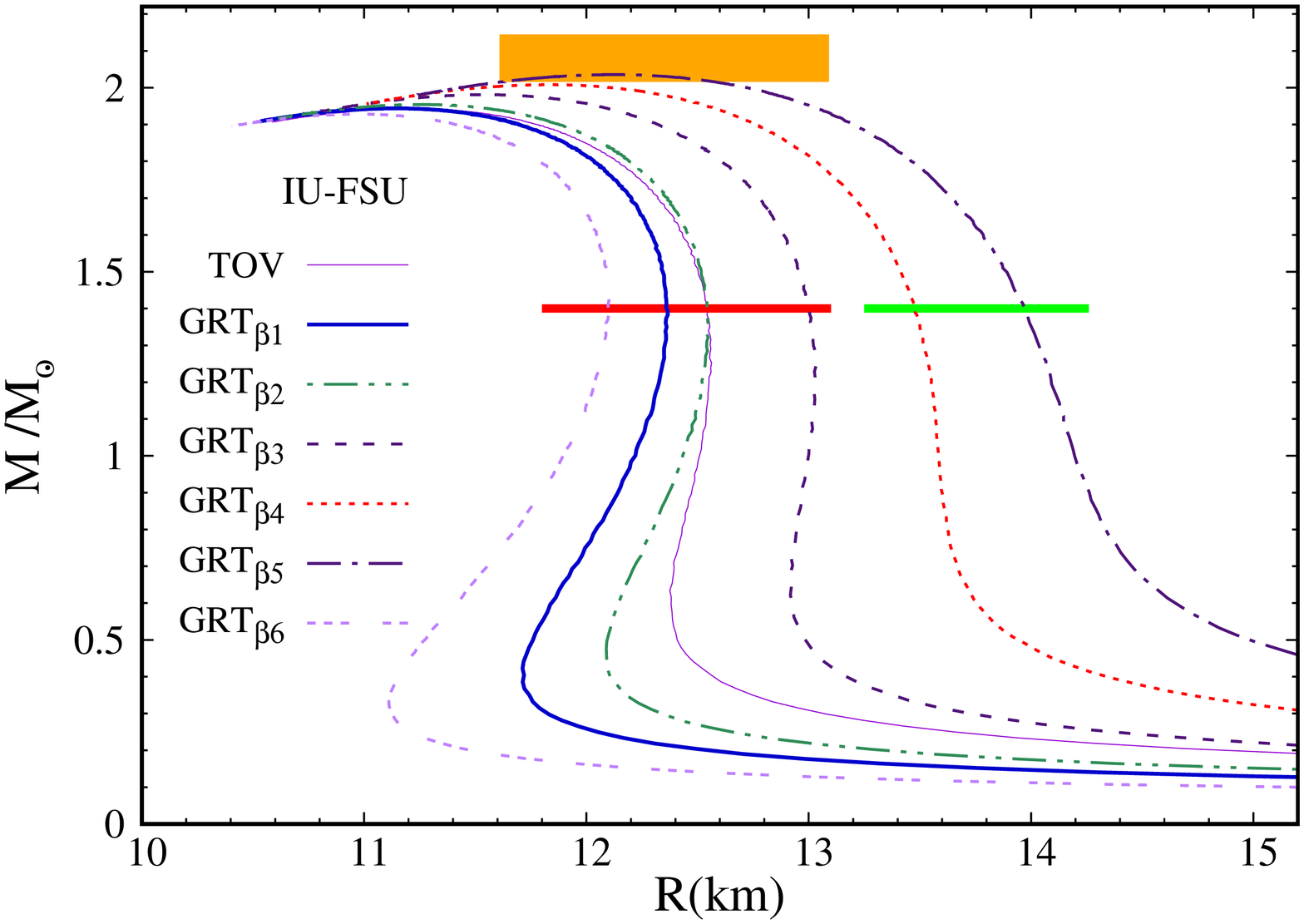} 
	\end{tabular}
	\caption{Mass-radius relation for a family of hadronic stars described with the IU-FSU EoS. We analyze the effects of varying the parameter $\alpha$ (top) while keeping the parameter $\beta$ fixed and the effects of varying the parameter $\beta$ (bottom) while keeping $\alpha$ fixed. The red and green line segments correspond respectively to the radius range of the $1.4 M_\odot$ NS for PSR J0030 + 0451 and PREX-2. The orange rectangular region represents the interval of radius estimates for the $2.08 \pm 0.07 M_\odot$ NS PSR J0740+6620.}
	\label{fig_IUFSU2}
\end{figure}

\begin{figure}[h]
	\centering
	\begin{tabular}{ll}
		\includegraphics[width=5.8cm,angle=270]{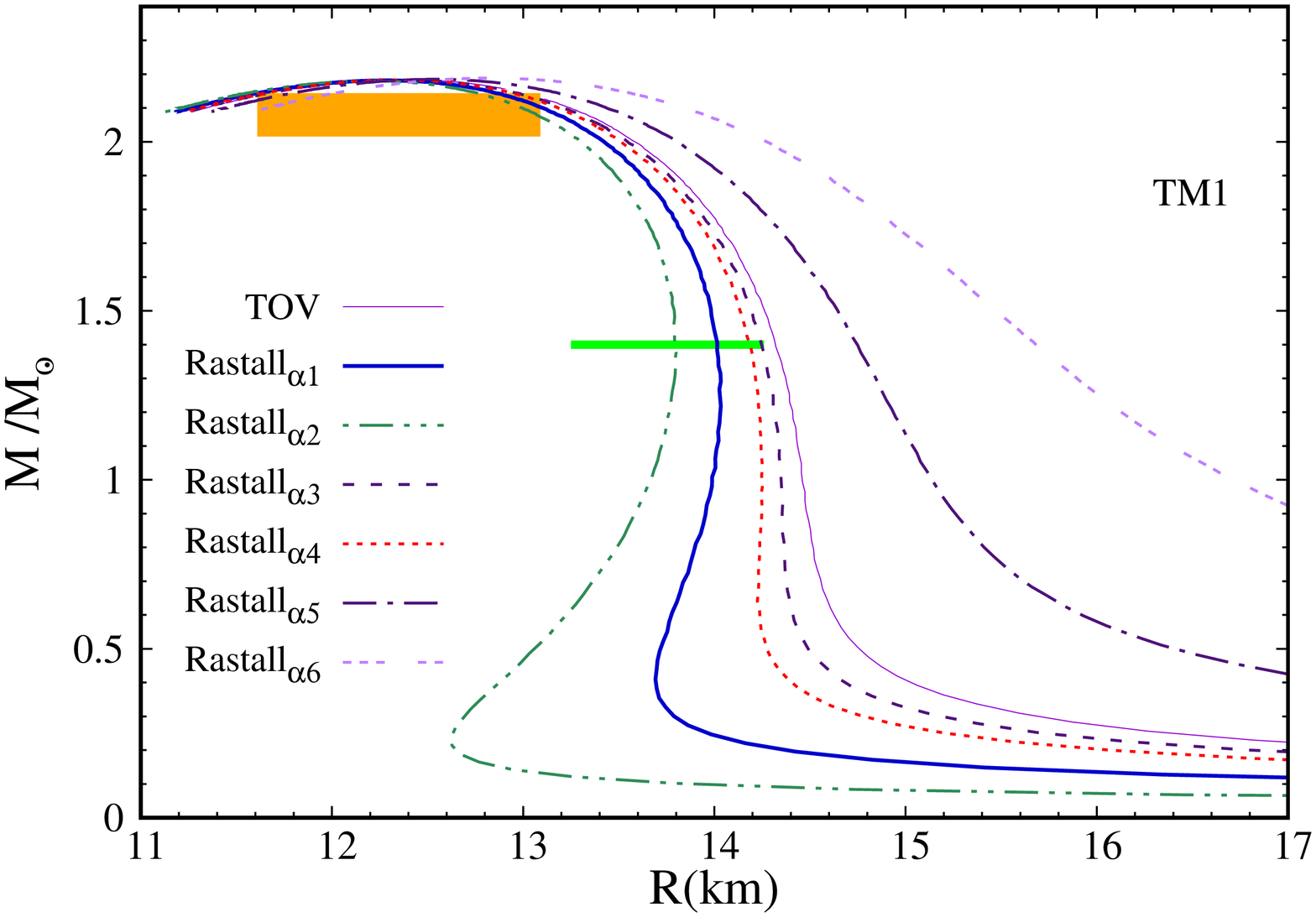}\\
		\includegraphics[width=5.8cm,angle=270]{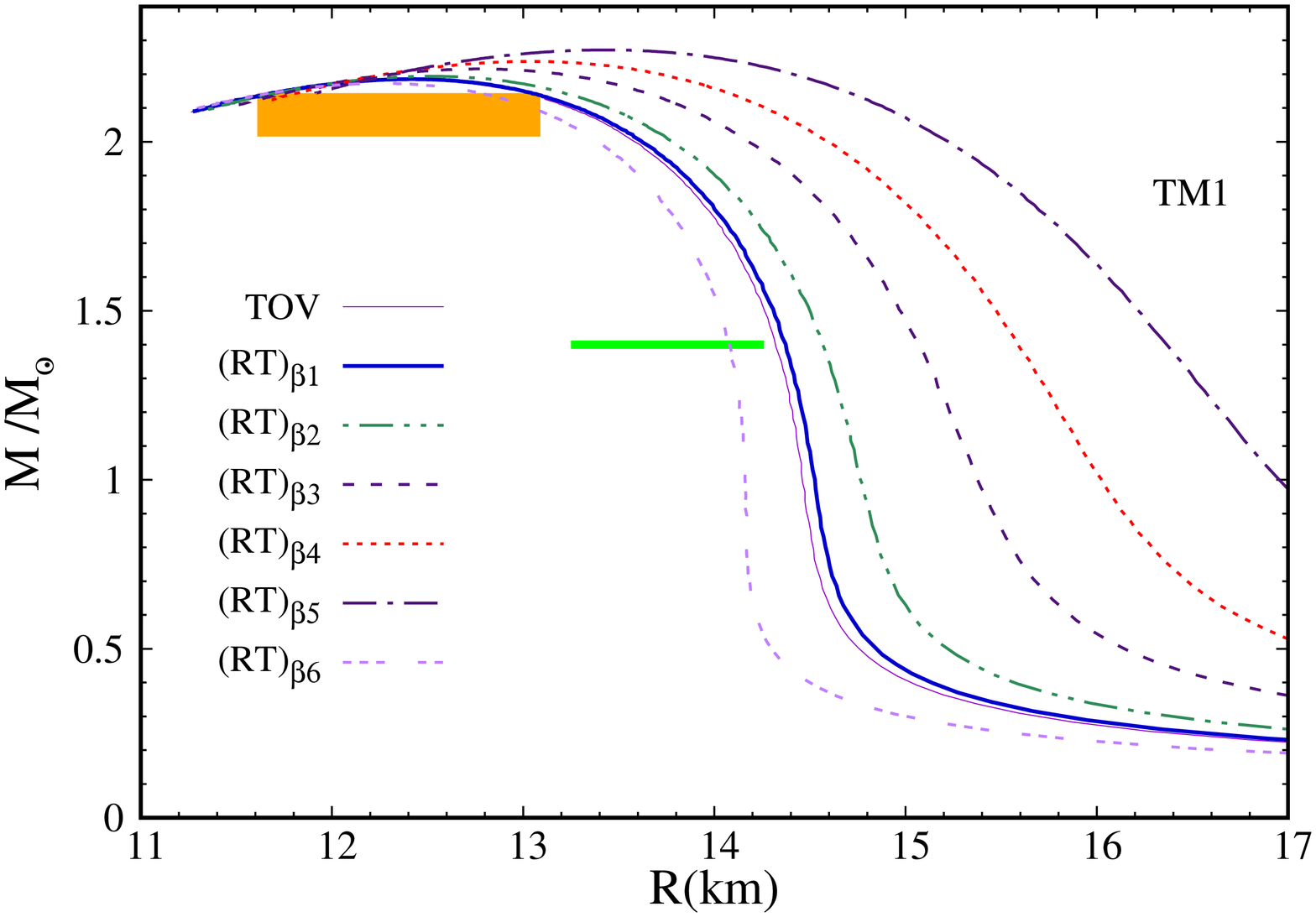} 
	\end{tabular}
	\caption{Mass-radius relation for a family of hadronic stars described with the TM1 EoS. We analyze the effects caused by varying the parameter $\alpha$ (top) while keeping the parameter $\beta$ null and the effects of varying the parameter $\beta$ (bottom) while keeping $\alpha$ null. The green line segment and orange rectangular region correspond respectively to the radius range for the $1.4 M_\odot$ NS PREX-2 and the $2.08 \pm 0.07 M_\odot$ NS PSR J0740+6620.}
	\label{fig_TM11}
\end{figure}

\begin{figure}[h]
	\centering
	\begin{tabular}{ll}
		\includegraphics[width=5.8cm,angle=270]{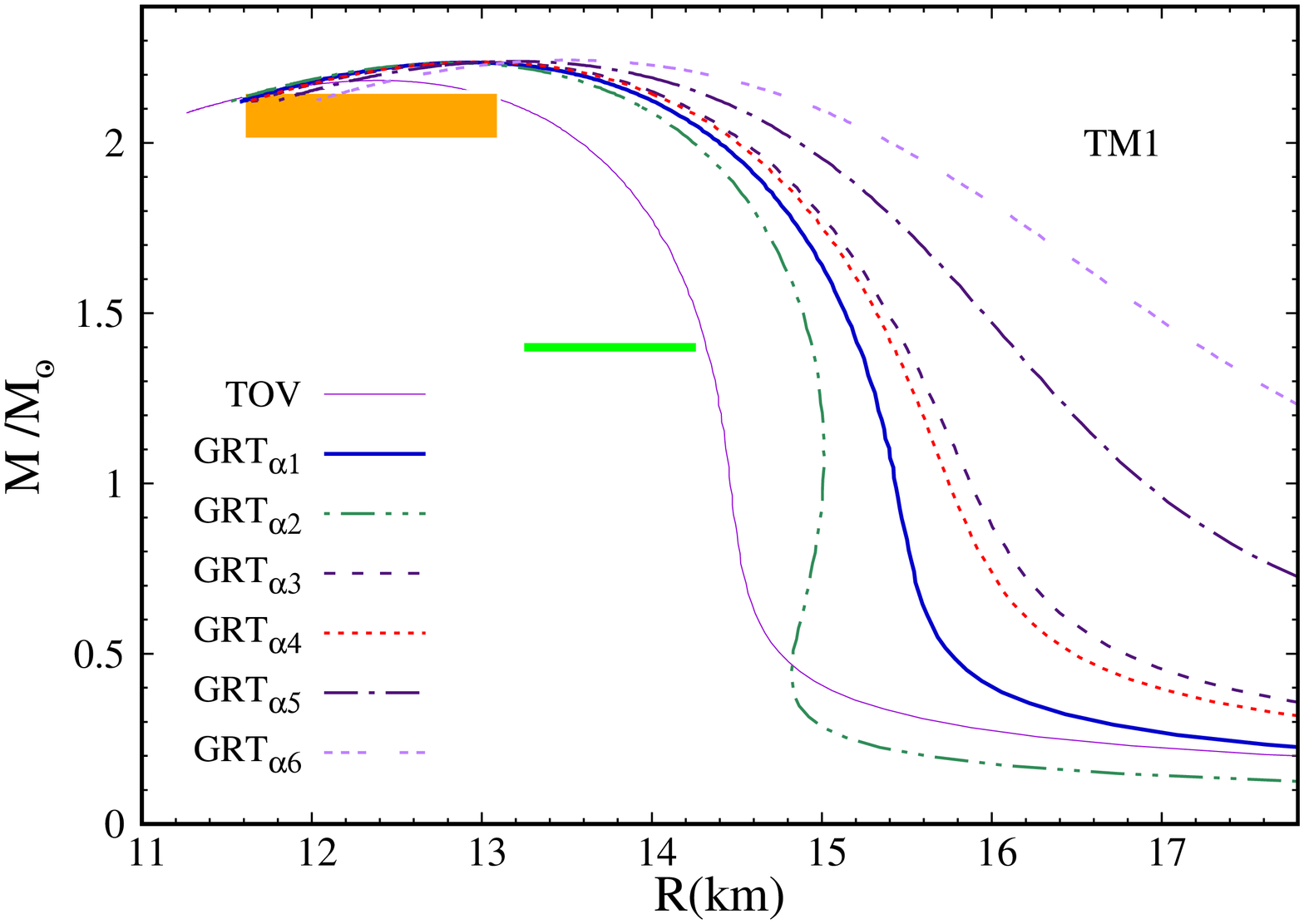}\\
		\includegraphics[width=5.8cm,angle=270]{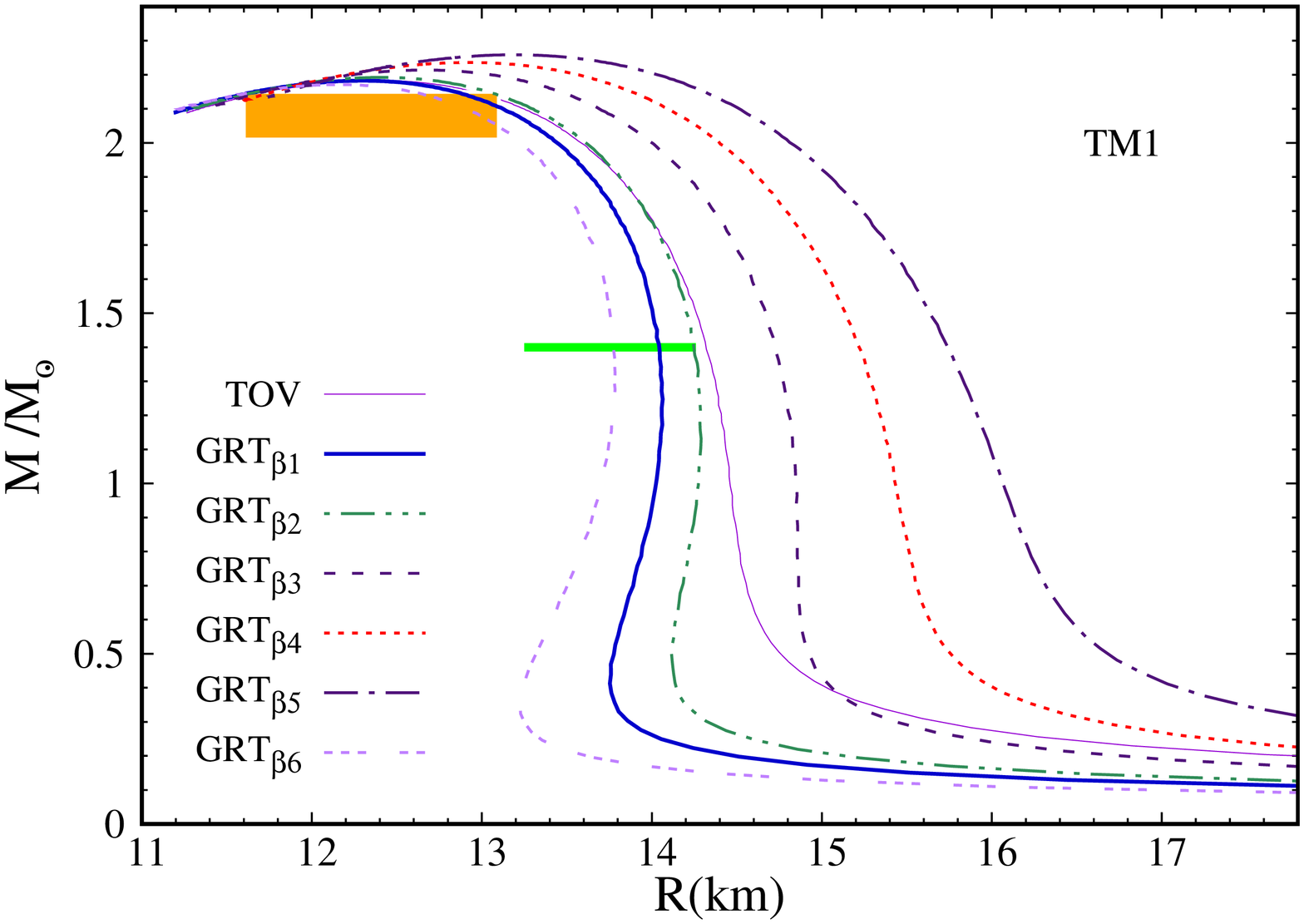} 
	\end{tabular}
	\caption{Mass-radius relation for a family NS within the TM1 EoS. We analyze the effects caused by varying the parameter $\alpha$ (top) while keeping the parameter $\beta$ fixed and the effects of varying the parameter $\beta$ (bottom) while keeping $\alpha$ fixed. The green line segment and orange rectangular region correspond respectively to the radius range for the $1.4 M_\odot$ NS PREX-2 and the $2.08 \pm 0.07 M_\odot$ NS PSR J0740+6620.}
	\label{fig_TM12}
\end{figure}

The differential equations (\ref{mass}) and (\ref{plinha}) for the stellar structure can be integrated numerically for the three unknown functions $m$, $p$ and $\rho$. Note that this integration occurs from the center to its surface, which is characterized by a point where $p$ vanishes. From different values of the EoS input
central density, $\rho_{c}$, and from the generalized Rastall's parameters,  $\alpha$ and $\beta$, we construct the macroscopic properties, i.e., the values of mass and corresponding radius for a family of neutron stars. The results are shown below in the tables and in the corresponding figures (mass-radius profiles). In Figs. (\ref{fig_IUFSU1}) and (\ref{fig_TM11}) we study the effect of Rastall (top) and the additional $RT$ term (bottom) alone, while in Figs. (\ref{fig_IUFSU2}) and (\ref{fig_TM12}) we analyse the effect of the generalized Rastall gravity with both terms included. The solutions for the standard case of GR are obtained numerically by using $\alpha = 0$ and $\beta = 0$. They are represented by continuous purple lines in the figures and the resulting values for the maximum mass and the corresponding radius for this solution are listed in the Tables.

Concerning astrophysical applications, we compare our results with recent observations and experiments. According to \cite{Cromartie}, the millisecond pulsars are among the most useful astrophysical objects in the Universe for testing fundamental physics, because they impose some of the most stringent constraints on high-density nuclear physics in the stellar interior. 
	Recently, the NICER mission reported pulsar observations for canonical ($1.4~M_\odot$) and massive ($2.0~M_\odot$) NS. These measurements provide a constraint of 11.80 km $\leq R_{1.4} \leq$ 13.10 km for the $1.4~ M_\odot$ NS PSR J0030+0451 \cite{Miller_2019} (horizontal line segment in red color shown in Figures \ref{fig_IUFSU1} and \ref{fig_IUFSU2}) and 11.60 km $\leq R \leq$ 13.10 km for a NS mass in the range $2.01~  M_\odot \leq M \leq 2.15~  M_\odot$ PSR J0740+6620 \cite{Miller2021} (the rectangular region in orange color shown in  Figures \ref{fig_IUFSU1} to \ref{fig_TM12}). From the nuclear physics point of view, the authors of Ref. \cite{PREX2021} used the recent measurement of neutron skin on $^{208}$Pb by PREX-2 to constrain the density dependence of the symmetry energy of the EoS. Remarkably, one can connect this constraint to NS radius predictions. In Ref. \cite{PREX2021} the connection of terrestrial experiments with astrophysical observations leads to a prediction of the radius of the canonical $1.4 ~M_\odot$ of 13.25 km $\lesssim R_{1.4} \lesssim$ 14.26 km. We confront our results with this constraint using a horizontal line segment in green color shown in all Figures \ref{fig_IUFSU1} to \ref{fig_TM12}.

%%%%%%%%%%%%%%%%%%%%%%%%%%%%%%%%%%%%%%%%%%%%%%%%%%%%[h!]
%\FloatBarrier

%%%%%%%%%%%%%%%%%%%%%%%%%%%%%%%%%%%%%%%%%%%%%%%%%%%%
%\FloatBarrier
%%%%%%%%%%%%%%%%%%%%%%%%%%%%%%%%%%%%%%%%%%%%%%%%%%%%

In Fig. \ref{fig_IUFSU1}, the effects of both parameters appearing in the generalized Rastall's gravity are individually analyzed for the IU-FSU EoS. In the top panel, we have the results corresponding to the Rastall's gravity solution, {\it i.e.}, $\alpha \neq 0$ and $\beta = 0$. The maximum mass is hardly affected by changing Rastall's parameter alone.
The radius of the
canonical NS ($M = 1.4 M_{\odot} $) is considerably affected. Note a bigger (smaller) radius for the most negative (positive) values of $\alpha$ within the range shown in Table I.
In the bottom panel of Fig. \ref{fig_IUFSU1}, we investigate the effects of the $RT$ term alone, {\it i.e.}, $\alpha=0$ and $\beta \neq 0$. Differently from the previous one, this parameter has a small, but non negligible effect on the maximum mass.
As $\beta$ grows we 
are able to reproduce more massive neutron stars, without introducing the parameter $\alpha$.
In this case, 
the radius of the canonical NS  increases with the increase of the $\beta$ parameter. These results show that while Rastall's theory alone affects basically the NS radius, the $RT$ term proposed in this work has an influence on the whole NS profiles.
It is also important to remember the effects of the Rastall theory on the radii of the canonical star, which are displaced with respect to the ones obtained with the TOV equations.

In Fig. \ref{fig_IUFSU2} and Table II, we
analyze the effect of having both parameters different from zero, still for the IU-FSU EoS. On the top panel, we fix $\beta = 5 \times 10^{-4}$  and vary $\alpha$. Note that the presence of a fixed $RT$ term slightly increases the
NS maximum mass, however, as in Fig. \ref{fig_IUFSU1}, the variation of $\alpha$ hardly affects the maximum mass.
The importance of the Rastall parameter is clearly seen on the NS radius of the whole family of stars. As in the previous case, the most positive $\alpha$ gives the smaller radius.
On the bottom panel, we fix 
$\alpha = 1 \times 10^{-5}$ and vary $\beta$.
We obtain an increase (decrease) of both NS radius and maximum mass for the most positive (negative) parameter. Here we call attention to the curve obtained with $\beta_1$ that, while keeping the same maximum mass as the TOV solution,
slightly decreases the 1.4 $M_{\odot}$ radius. The resulting values lie inside the range of the recent work \cite{CapanoNature}, where by combining data from multi-messenger observations and nuclear physics, the authors obtained the most stringent constraint to the canonical neutron star radius, $R_{1.4} = 11.0^{+0.9}_{-0.6}$ km. Note that the TOV solution for IU-FSU EoS lies in the lower band of the PSR J0030 (red line) constraint and fails to reproduce the PSR J040 (orange) and PREX-2 (green) constraints. The GRT solution impacts the mass-radius curves in such a way that with a more negative $\alpha$ the lower band of PSR J040 is achieved. Also, for $\alpha_1$ to $\alpha_5$, the canonical NS radius is outside the PSR J0030 (red) and inside the PREX-2 (green) constraint. When we keep $\alpha$ fixed and vary $\beta$ in the bottom panel of Fig. \ref{fig_IUFSU2} we note most of solutions inside the red constraint of PSR J0030 with $\beta_5$ respecting PSR J0740 and PREX-2. In this case, our modifications on $\beta$ increase the radius enough to be inside PREX-2 constraint. It is obvious that the constraints represented by the red and green segments cannot be both satisfied.

Recent results for neutron stars in the context of Rastall's theory can be found in Ref. \cite{black2,santos8}. We have checked that although the Rastall's gravity alone affects very little the maximum stellar mass, it considerably increases the corresponding radius, while the canonical radius of the star also increases \cite{black2}. On the other hand, the authors in \cite{santos8} have shown that it is possible to cause the maximum stellar mass to increase at the same time that the canonical radius of the star decreases, however, at the expense of adding a parameter of a second theory. In contrast, we show in our results that regardless of the two EoS tested, we reproduce results similar to those present in \cite{santos8}, however, within the same theory of gravity.

In Figs. \ref{fig_TM11} and \ref{fig_TM12} we repeat the previous analyses with a stiffer EoS, TM1. The results  obtained with various parameter values are displayed in Tables III and IV. The general effects of the parameters $\alpha$ and $\beta$ are the same as the ones obtained with the IU-FSU EoS. Note that the TOV solution for TM1 predicts a bigger radius for the canonical NS as compared with the constraints from NICER and PREX-2, a direct effect of the stiffness of this EoS. For this reason we have removed the PSR J0030 (red line on Figs. \ref{fig_IUFSU1} and \ref{fig_IUFSU2}) since the modifications of GRT are also outside of the range imposed by this constraint. We note that all curves with this EoS, GR or GRT, achieve high NS masses. We also note in Fig. \ref{fig_TM12} that for small positive and negative values of the parameter $\beta$ GRT can predict NS radius inside the PREX-2 constraint with negligible modifications on the maximum masses.

One should notice that TM1 shows a bigger canonical radius than the IU-FSU. However, with appropriate parameters, the generalized Rastall's gravity gives a slightly smaller radius for the NS of $M = 1.4 M_{\odot}$ while keeping the maximum mass above the required 2$M_{\odot}$.

It is important to remark that the use of the generalized Rastall's theory yields similar variations of the macroscopic quantities (as compared with the used of the TOV equations) which are  independent of the chosen nuclear EoS (within the two representative ones analysed in this work). Moreover, although the generalized version of the Rastall's gravity allows more flexibility in the calculation of the macroscopic stellar properties due to the inclusion of two independent parameters, it does not fix existing caveats of the EoS. Hence, an EoS that satisfies bulk nuclear matter properties is still required as input to the generalized Rastall's equations.

%%%%%%%%%%%%%%%%%%%%%%%%%%%%%%%%%%%%%%%%%%%%%%%%%%%%
%\FloatBarrier
%%%%%%%%%%%%%%%%%%%%%%%%%%%%%%%%%%%%%%%%%%%%%%%%%%%%
%%%%%%%%%%%%%%%%%%%%%%%%%%%%%%%%%%%%%%%%%%%%%%%%%%

\begin{figure}[h]
	\centering
		\includegraphics[width=5.8cm,angle=270]{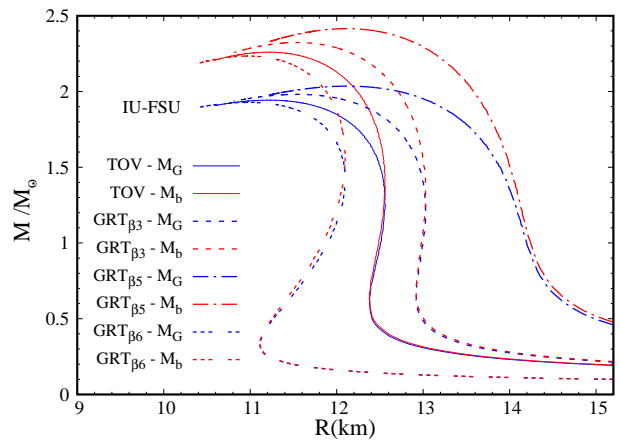}
		\caption{Comparison of gravitational mass (blue) and baryon mass (red) for the TOV solutions (continuous lines) and several Generalized Rastall (dashed).}
	\label{fig_mass}
\end{figure}

Lastly, we analyse whether the results obtained in this theory provide stars that are gravitationally bound which, according to \cite{glendenning,haensel2007neutron}, occurs when the gravitational mass $M_G$ is smaller than the rest mass $M_b$ (baryonic mass). We have computed the two masses and compared their values for all mass-radius solutions calculated in the present work and we add some typical results in Fig. \ref{fig_mass}, as an example. The figure shows the TOV solutions in continuous lines and Generalized Rastall in dashed lines. Blue lines show the gravitational mass, while the red ones show baryon mass. We can see that $M_b$ is always bigger than $M_G$. 
Note that the region where the blue and red lines are close refer to very low mass NS where stable solutions may not exist. The lightest NS ever measured has 1.17 $M_{\odot}$ \cite{Martinez2015}.

%%%%%%%%%%%%%%%%%%%%%%%%%%%%%%%%%%%%%%%%%%%%%%%%%%%%
%%%%%%%%%%%%%%%%%%%%%%%%%%%%%%%%%%%%%%%%%%%%%%%%%%

%\newpage
\section{Conclusions} \label{3}

In this work we have generalized Rastall's theory of gravity. Original Rastall's gravity breaks the energy-momentum conservation making $T_{\: \:\:\nu;\:\mu}^{\mu}
= A_{\nu} =
\lambda R_{,\:\nu}$, where a dependence on the curvature $R$
appears on the derivative of $T_{\: \:\:\nu;\:\mu}^{\mu}$.
We propose that this derivative also depends on the trace of the energy-momentum tensor, $T$, {\it i.e.} the function $A_{\nu}$ is given now by
$A_{\nu}=(\alpha \delta_{\:\:\nu}^{\mu}R + \beta \delta_{\:\:\nu}^{\mu} RT)_{;\:\mu}$. Initially we have discussed the external solution in the case of a space-time with spherical symmetry. We have pointed out that the solution with $R=0$ represents the gravitational field outside a spherical mass in our system, and unlike the GR, the vacuum solution in the context of the generalized Rastall's gravity gives two types of space-time solutions, depending on the choice of the trace of the energy-momentum tensor. This property of the external solution can be regarded as an intrinsic feature of the generalized theory since other choices for breaking the conservation law lead to the same result. In addition,  we have carefully studied the Newtonian limit of the theory and shown that the gravitational parameter $k$ depends on the parameter $\alpha$ in contrast to the constant value $k= 8 \pi G$ in general relativity and the parameter $\beta$ must be negligible in the case of a weak gravitational field.

We tested the theory in neutron stars using two different RMF EoS as inputs and noted a considerable effect of the alternative gravity theory in the NS mass-radius diagrams.
The results presented here show that, with small deviations from the GR case, an important change on the NS profile can be obtained within the same nuclear physics inputs. 

The recent result of LIGO-Virgo \cite{NS_2p6MsunLIGO} with a possible NS of 2.6$M_{\odot}$ is also of particular interest. If such massive NS is confirmed by
future observations it will be a big challenge for the compact objects community to describe it. From the nuclear physics point of view, one needs a very stiff EoS at high densities to support such a high maximum mass together with a soft EoS at low densities to keep a radius of the order of 11 km for the canonical 1.4$M_{\odot}$ NS. From the gravity point of view one can explore theories beyond GR such as the one examined in the present paper. Our work generalizes the results found in \cite{black2} and, furthermore we take into account the possibility of more complex combinations of matter and geometry that can be associated to realistic configurations of the matter inside NS. The generalization opens a new line of research in the context of non-conservative gravitational theories in which various choices of the conservation law can be tested in different astrophysical and cosmological problems.

In future works, it would be interesting to study gravitational effects in astrophysical and cosmological systems due to the choice of other combinations of $R$ and $T$ and/or to investigate further effects of the interplay between non-conservative gravitational theories and Planck scale phenomenological approaches \cite{santos8}. Moreover, the application of the present formalism in anisotropic stars is the next step towards a more realistic description of these compact objects. 

At the formal level, it would be of interest to the community the search for a Lagrangian for Rastall and Generalized Rastall formalisms, which could allow one to deal with other aspects of field theory, like investigation about the presence of ghosts. Regarding the post-Newtonian corrections, it is worth mentioning that theories with nonzero divergence of the energy-momentum tensor which satisfy certain consistency conditions can be subject to experimental verification in the post-Newtonian expansions \cite{posnewton1}. A detailed analysis of post-Newtonian terms in the context of Rastall's gravity can be found in \cite{posnewton2}. We believe that in future works, a similar study can be done with the help of the present generalized version of Rastall's gravity.

%%%%%%%%%%%%%%%%%%%%%%%%%%%%%%%%%%%%%%%%%%%%%%%%%%%%%%%%%%%%%%%%%%%%%%%%%%%%%%%%%%%%%%%%%%%%%%

\begin{table*}[t]
	\addtolength{\tabcolsep}{-2.5pt}
	\caption{Macroscopic properties for different values of the $\alpha$ and $\beta$ parameters corresponding to the mass-radius diagram in FIG.\ref{fig_IUFSU1}. The quantities $M_{max}$, $R_{1.4}$, $\Bar{\rho}_{c}$, $C_{max}$ and $C_{1.4}$ correspond respectively to maximum mass, canonical star radius, central energy density and compactness. $\alpha$ is dimensionless and $\beta$ has dimensions $fm^4$.}
	{\footnotesize
		\begin{tabular}{c|c|c|ccccccc}
			\hline
			\hline
			&  & \ \ TOV \ \ & Rastall$_{\alpha1}$ \ \ \ \ & Rastall$_{\alpha2}$ \ \ \ \ & Rastall$_{\alpha3}$ \ \ & Rastall$_{\alpha4}$  \ \ & Rastall$_{\alpha5}$ \ \ & Rastall$_{\alpha6}$ \tabularnewline 
			
			\hline
			& $\boldsymbol{\alpha}$  &  \textbf{0.0}  & $1 \times 10^{-5}$ & $2 \times 10^{-5}$ & $2 \times 10^{-6}$  &  $4 \times 10^{-6}$  & $-1 \times 10^{-5}$  & $-3 \times 10^{-5}$  \tabularnewline
			
			Parameters & $\beta$  &  0.0  &  0.0 &  0.0 &  0.0  & 0.0  & 0.0 & 0.0  \tabularnewline
			
			\hline
			& $M_{max}$  &  1.942 $M_\odot$ &  1.941 $M_\odot$  & 1.939 $M_\odot$  &  1.942 $M_\odot$  & 1.942 $M_\odot$  & 1.944 $M_\odot$ & 1.947 $M_\odot$\tabularnewline
			
			IU-FSU %& $R_{M_{max}}$ &  11.22 km  \ \ &  11.26 km   \ \ & 11.34 km \ \ & 11.59 km  \ \ & 12.23 km  \ \ & 11.09 km \ \ & 10.14 km  \tabularnewline
			
			&$R_{1.4}$& 12.545 km  \ \ &  12.315 km \ \ & 12.142 km  \ \ & 12.488 km  \ \ & 12.442 km  \ \ & 12.850 km \ \ & 13.524 km  \tabularnewline
			
			&$\Bar{\rho}_{c}$& 6.365 $fm^{-4}$  \ \ &  6.366 $fm^{-4}$ \ \ & 6.389 $fm^{-4}$  \ \ & 6.364 $fm^{-4}$  \ \ & 6.365 $fm^{-4}$  \ \ & 6.343 $fm^{-4}$ \ \ & 6.317 $fm^{-4}$  \tabularnewline
			
			&$C_{max}$& 0.173 \ \ &  0.174  \ \ & 0.175 \ \ & 0.173  \ \ & 0.173  \ \ & 0.171  \ \ & 0.167  \tabularnewline
			
			&$C_{1.4}$& 0.111 \ \ & 0.113  \ \ & 0.115  \ \ & 0.112  \ \ & 0.112  \ \ & 0.108  \ \ & 0.103  \tabularnewline
			\hline
			&  &   \ \ &    \ \ &\ \ &   \ \ &    \ \  &   \ \ &  \ \ \tabularnewline
			
			&  &  General  \ \ &    \ \ & \ \ & \ \ & Generalized   \ \  &   \ \ &  \ \ \tabularnewline
			&  &  Relativity  \ \ &    \ \ & \ \ & \ \ & Rastall's gravity \ \  &   \ \ &  \ \ \tabularnewline
			&  &    \ \ &     \ \ & \ \ &   \ \ &    \ \  &   \ \ &  \ \ \tabularnewline
			
			\hline
			&  & \ \ TOV \ \ & (RT)$_{\beta1}$ \ \ & (RT)$_{\beta2}$ \ \ & (RT)$_{\beta3}$ \ \ & (RT)$_{\beta4}$ \ \ & (RT)$_{\beta5}$  \ \ & (RT)$_{\beta6}$ \tabularnewline 
			\hline
			& $\boldsymbol{\beta}$  &  \textbf{0.0}  &  $2 \times 10^{-5}$ &  $1 \times 10^{-4}$  &  $3 \times 10^{-4}$  & $5 \times 10^{-4}$ & $7 \times 10^{-4}$  & $-1 \times 10^{-4}$ \tabularnewline
			
			Parameters & $\alpha$  &  0.0  & 0.0 & 0.0 & 0.0  & 0.0 &  0.0  & 0.0  \tabularnewline
			\hline
			& $M_{max}$  &  1.942 $M_\odot$ &  1.945 $M_\odot$  & 1.955 $M_\odot$  &  1.982 $M_\odot$  &  2.009 $M_\odot$ & 2.037 $M_\odot$  & 1.929 $M_\odot$\tabularnewline
			
			IU-FSU %& $R_{M_{max}}$ &  11.22 km  \ \ &  11.28 km   \ \ & 11.61 km \ \ & 12.85 km  \ \ & 13.34 km  \ \  & 13.94 km  \ \ & 11.18 km \ \ \tabularnewline
			
			&$R_{1.4}$& 12.545 km  \ \ &  12.589 km \ \ & 12.778 km  \ \ & 13.248 km  \ \ & 13.752 km  \ \ & 14.270 km  \ \ & 12.312 km \ \ \tabularnewline
			
			&$\Bar{\rho}_{c}$& 6.365 $fm^{-4}$  \ \ & 6.346 $fm^{-4}$  \ \ &  6.251 $fm^{-4}$ \ \ & 6.022 $fm^{-4}$  \ \ & 5.812 $fm^{-4}$  \ \ & 5.621 $fm^{-4}$  \ \ & 6.480 $fm^{-4}$ \ \ \tabularnewline
			
			&$C_{max}$& 0.173 \ \ &  0.172  \ \ & 0.172 \ \ & 0.170  \ \ & 0.168  \ \ & 0.166  \ \ & 0.174  \tabularnewline
			
			&$C_{1.4}$& 0.111 \ \ & 0.111  \ \ & 0.109  \ \ & 0.105  \ \ & 0.101  \ \ & 0.098  \ \ & 0.113  \tabularnewline
			\hline
			\hline
		\end{tabular}
	}
\end{table*}

%%%%%%%%%%%%%%%%%%%%%%%%%%%%%%%%%%%%%%%%%%%%%%%%%%%%%%%%%%%%%%%%%%%%%%%%%%%%%%%%%%%%%%%%%%%%%%

\begin{table*}[t]
	\addtolength{\tabcolsep}{-2.5pt}
	\caption{Macroscopic properties for different values of the $\alpha$ and $\beta$ parameters corresponding to the mass-radius diagram in FIG.\ref{fig_IUFSU2}. The quantities $M_{max}$, $R_{1.4}$, $\Bar{\rho}_{c}$, $C_{max}$ and $C_{1.4}$ correspond respectively to maximum mass, canonical star radius, central energy density and compactness. $\alpha$ is dimensionless and $\beta$ has dimensions $fm^4$.}
	{\footnotesize
		\begin{tabular}{c|c|c|ccccccc}
			\hline
			\hline
			&  & \ \ TOV \ \ & GRT$_{\alpha1}$ \ \ \ \ & GRT$_{\alpha2}$ \ \ \ \ & GRT$_{\alpha3}$ \ \ & GRT$_{\alpha4}$ \ \ & GRT$_{\alpha5}$ \ \ & GRT$_{\alpha6}$ \tabularnewline 
			
			\hline
			& $\boldsymbol{\alpha}$  &  \textbf{0.0}  & $1 \times 10^{-5}$ & $2 \times 10^{-5}$ & $2 \times 10^{-6}$  &  $4 \times 10^{-6}$  & $-1 \times 10^{-5}$  & $-3 \times 10^{-5}$  \tabularnewline
			
			Parameters & $\beta$  &  0.0  &  $5 \times 10^{-4}$ &  $5 \times 10^{-4}$ & $5 \times 10^{-4}$  & $5 \times 10^{-4}$  & $5 \times 10^{-4}$ & $5 \times 10^{-4}$  \tabularnewline
			
			\hline
			& $M_{max}$  &  1.942 $M_\odot$ &  2.007 $M_\odot$  & 2.006 $M_\odot$  &  2.009 $M_\odot$  & 2.008 $M_\odot$  & 2.011 $M_\odot$ & 2.014 $M_\odot$\tabularnewline
			
			IU-FSU %& $R_{M_{max}}$ &  11.22 km  \ \ &  11.64 km   \ \ & 11.70 km \ \ & 11.90 km  \ \ & 12.44 km  \ \ & 11.52 km \ \ & 10.87 km  \tabularnewline
			
			&$R_{1.4}$& 12.545 km  \ \ &  13.480 km \ \ & 13.267 km  \ \ & 13.686 km  \ \ & 13.629 km \ \ & 14.141 km \ \ & 14.978 km  \tabularnewline
			
			&$\Bar{\rho}_{c}$& 6.365 $fm^{-4}$ \ \ & 5.833 $fm^{-4}$ \ \ & 5.836 $fm^{-4}$ \ \ & 5.812 $fm^{-4}$ \ \ & 5.832 $fm^{-4}$ \ \ & 5.808 $fm^{-4}$ \ \ & 5.783 $fm^{-4}$ \tabularnewline
			
			&$C_{max}$& 0.173 \ \ &  0.169  \ \ & 0.170 \ \ & 0.168  \ \ & 0.168  \ \ & 0.166  \ \ & 0.162  \tabularnewline
			
			&$C_{1.4}$& 0.111 \ \ & 0.103  \ \ & 0.105  \ \ & 0.102  \ \ & 0.102  \ \ & 0.099  \ \ & 0.093  \tabularnewline
			\hline
			&  &   \ \ &    \ \ &\ \ &   \ \ &    \ \  &   \ \ &  \ \ \tabularnewline
			
			&  &  General  \ \ &    \ \ &\ \ &   \ \ & Generalized    \ \  &   \ \ &  \ \ \tabularnewline
			&  &  Relativity  \ \ &    \ \ &\ \ &   \ \ & Rastall's gravity \ \  &   \ \ &  \ \ \tabularnewline
			&  &    \ \ &     \ \ &\ \ &   \ \ &    \ \  &   \ \ &  \ \ \tabularnewline
			
			\hline
			&  & \ \ TOV \ \ & GRT$_{\beta1}$ \ \ & GRT$_{\beta2}$ \ \ & GRT$_{\beta3}$ \ \ & GRT$_{\beta4}$ \ \ & GRT$_{\beta5}$ \ \ & GRT$_{\beta6}$ \tabularnewline 
			\hline
			& $\boldsymbol{\beta}$  &  \textbf{0.0}  &  $2 \times 10^{-5}$ &  $1 \times 10^{-4}$  &  $3 \times 10^{-4}$  & $5 \times 10^{-4}$ & $7 \times 10^{-4}$  & $-1 \times 10^{-4}$ \tabularnewline
			
			Parameters & $\alpha$  &  0.0  & $1 \times 10^{-5}$ & $1 \times 10^{-5}$ & $1 \times 10^{-5}$  & $1 \times 10^{-5}$ &  $1 \times 10^{-5}$  & $1 \times 10^{-5}$  \tabularnewline
			\hline
			& $M_{max}$  &  1.942 $M_\odot$ &  1.943 $M_\odot$  & 1.954 $M_\odot$  &  1.980 $M_\odot$  &  2.007 $M_\odot$ & 2.035 $M_\odot$  & 1.928 $M_\odot$\tabularnewline
			
			IU-FSU %& $R_{M_{max}}$ &  11.22 km  \ \ &  10.09 km   \ \ & 10.87 km \ \ & 12.54 km  \ \ & 13.07 km  \ \  & 13.67 km  \ \ & 9.82 km \ \ \tabularnewline
			
			&$R_{1.4}$& 12.545 km  \ \ &  12.358 km \ \ & 12.544 km  \ \ & 13.004 km  \ \ & 13.474 km  \ \ & 13.970 km  \ \ & 12.094 km \ \ \tabularnewline
			
			&$\Bar{\rho}_{c}$& 6.365 $fm^{-4}$ \ \ & 6.347 $fm^{-4}$ \ \ & 6.252 $fm^{-4}$ \ \ & 6.043 $fm^{-4}$ \ \ & 5.833 $fm^{-4}$ \ \ & 5.624 $fm^{-4}$ \ \ & 6.481 $fm^{-4}$ \tabularnewline
			
			&$C_{max}$& 0.173 \ \ &  0.174  \ \ & 0.173 \ \ & 0.171  \ \ & 0.169  \ \ & 0.167  \ \ & 0.175  \tabularnewline
			
			&$C_{1.4}$& 0.111 \ \ & 0.113  \ \ & 0.111  \ \ & 0.107  \ \ & 0.103  \ \ & 0.100  \ \ & 0.115  \tabularnewline
			\hline
			\hline
		\end{tabular}
	}
\end{table*}
%%%%%%%%%%%%%%%%%%%%%%%%%%%%%%%%%%%%%%%%%%%%%%%%%%%%%%%%%%%%%%%%%%%%%%%%%%%%%%%%%%%%%%%%%%%%%%

\begin{table*}[t]
	\addtolength{\tabcolsep}{-2.5pt}
	\caption{Macroscopic properties for different values of the $\alpha$ and $\beta$ parameters corresponding to the mass-radius diagram in FIG.\ref{fig_TM11}. The quantities $M_{max}$, $R_{1.4}$, $\Bar{\rho}_{c}$, $C_{max}$ and $C_{1.4}$ correspond respectively to maximum mass, canonical star radius, central energy density and compactness. $\alpha$ is dimensionless and $\beta$ has dimensions $fm^4$.}
	{\footnotesize
		\begin{tabular}{c|c|c|ccccccc}
			\hline
			\hline
			&  & \ \ TOV \ \ & Rastall$_{\alpha1}$ \ \ \ \ & Rastall$_{\alpha2}$ \ \ \ \ & Rastall$_{\alpha3}$ \ \ & Rastall$_{\alpha4}$  \ \ & Rastall$_{\alpha5}$ \ \ & Rastall$_{\alpha6}$ \tabularnewline 
			
			\hline
			& $\boldsymbol{\alpha}$  &  \textbf{0.0}  & $1 \times 10^{-5}$ & $2 \times 10^{-5}$ & $2 \times 10^{-6}$  &  $4 \times 10^{-6}$  & $-1 \times 10^{-5}$  & $-3 \times 10^{-5}$  \tabularnewline
			
			Parameters & $\beta$  &  0.0  &  0.0 &  0.0 &  0.0  & 0.0  & 0.0 & 0.0  \tabularnewline
			
			\hline
			& $M_{max}$  &  2.183 $M_\odot$ &  2.181 $M_\odot$  & 2.180 $M_\odot$  &  2.183 $M_\odot$  & 2.182 $M_\odot$  & 2.185 $M_\odot$ & 2.188 $M_\odot$\tabularnewline
			
			TM1 %& $R_{M_{max}}$ &  11.22 km  \ \ &  11.26 km   \ \ & 11.34 km \ \ & 11.59 km  \ \ & 12.23 km  \ \ & 11.09 km \ \ & 10.14 km  \tabularnewline
			
			&$R_{1.4}$& 14.332 km  \ \ &  14.011 km \ \ & 13.797 km  \ \ & 14.249 km  \ \ & 14.185 km  \ \ & 14.748 km \ \ & 15.676 km  \tabularnewline
			
			&$\Bar{\rho}_{c}$& 5.347 $fm^{-4}$ \ \ & 5.351 $fm^{-4}$ \ \ & 5.380 $fm^{-4}$ \ \ & 5.349 $fm^{-4}$ \ \ & 5.350 $fm^{-4}$ \ \ & 5.344 $fm^{-4}$ \ \ & 5.313 $fm^{-4}$ \tabularnewline
			
			&$C_{max}$& 0.176 \ \ &  0.177  \ \ & 0.178 \ \ & 0.176  \ \ & 0.176  \ \ & 0.174  \ \ & 0.170  \tabularnewline
			
			&$C_{1.4}$& 0.097 \ \ & 0.099  \ \ & 0.101  \ \ & 0.098  \ \ & 0.098  \ \ & 0.094  \ \ & 0.089  \tabularnewline
			\hline
			&  &   \ \ &    \ \ &\ \ &   \ \ &    \ \  &   \ \ &  \ \ \tabularnewline
			
			&  &  General  \ \ &    \ \ & \ \ & \ \ & Generalized   \ \  &   \ \ &  \ \ \tabularnewline
			&  &  Relativity  \ \ &    \ \ & \ \ & \ \ & Rastall's gravity \ \  &   \ \ &  \ \ \tabularnewline
			&  &    \ \ &     \ \ & \ \ &   \ \ &    \ \  &   \ \ &  \ \ \tabularnewline
			
			\hline
			&  & \ \ TOV \ \ & (RT)$_{\beta1}$ \ \ & (RT)$_{\beta2}$ \ \ & (RT)$_{\beta3}$ \ \ & (RT)$_{\beta4}$ \ \ & (RT)$_{\beta5}$  \ \ & (RT)$_{\beta6}$ \tabularnewline 
			\hline
			& $\boldsymbol{\beta}$  &  \textbf{0.0}  &  $2 \times 10^{-5}$ &  $1 \times 10^{-4}$  &  $3 \times 10^{-4}$  & $5 \times 10^{-4}$ & $8 \times 10^{-4}$  & $-1 \times 10^{-4}$ \tabularnewline
			
			Parameters & $\alpha$  &  0.0  & 0.0 & 0.0 & 0.0  & 0.0 &  0.0  & 0.0  \tabularnewline
			\hline
			& $M_{max}$  &  2.183 $M_\odot$ &  2.185 $M_\odot$  & 2.194 $M_\odot$  &  2.215 $M_\odot$  &  2.237 $M_\odot$ & 2.271 $M_\odot$  & 2.172 $M_\odot$\tabularnewline
			
			TM1 %& $R_{M_{max}}$ &  11.22 km  \ \ &  11.28 km   \ \ & 11.61 km \ \ & 12.85 km  \ \ & 13.34 km  \ \  & 13.94 km  \ \ & 11.18 km \ \ \tabularnewline
			
			&$R_{1.4}$& 14.322 km  \ \ &  14.365 km \ \ & 14.568 km  \ \ & 15.067 km  \ \ & 15.595 km  \ \ & 16.366 km  \ \ & 14.075 km \ \ \tabularnewline
			
			&$\Bar{\rho}_{c}$& 5.347 $fm^{-4}$ \ \ & 5.348 $fm^{-4}$ \ \ & 5.267 $fm^{-4}$ \ \ & 5.133 $fm^{-4}$ \ \ & 4.974 $fm^{-4}$ \ \ & 4.760 $fm^{-4}$ \ \ & 5.427 $fm^{-4}$ \tabularnewline
			
			&$C_{max}$& 0.176 \ \ &  0.176  \ \ & 0.175 \ \ & 0.173  \ \ & 0.171  \ \ & 0.169  \ \ & 0.177  \tabularnewline
			
			&$C_{1.4}$& 0.097 \ \ & 0.097  \ \ & 0.096  \ \ & 0.092  \ \ & 0.089  \ \ & 0.085  \ \ & 0.099  \tabularnewline
			\hline
			\hline
		\end{tabular}
	}
\end{table*}

%%%%%%%%%%%%%%%%%%%%%%%%%%%%%%%%%%%%%%%%%%%%%%%%%%%%%%%%%%%%%%%%%%%%%%%%%%%%%%%%%%%%%%%%%%%%%%
\begin{table*}[t]
	\addtolength{\tabcolsep}{-2.5pt}
	\caption{Macroscopic properties for different values of the $\alpha$ and $\beta$ parameters corresponding to the mass-radius diagram in FIG.\ref{fig_TM12}. The quantities $M_{max}$, $R_{1.4}$, $\Bar{\rho}_{c}$, $C_{max}$ and $C_{1.4}$ correspond respectively to maximum mass, canonical star radius, central energy density and compactness. $\alpha$ is dimensionless and $\beta$ has dimensions $fm^4$.}
	{\footnotesize
		\begin{tabular}{c|c|c|ccccccc}
			\hline
			\hline
			&  & \ \ TOV \ \ & GRT$_{\alpha1}$ \ \ \ \ & GRT$_{\alpha2}$ \ \ \ \ & GRT$_{\alpha3}$ \ \ & GRT$_{\alpha4}$ \ \ & GRT$_{\alpha5}$ \ \ & GRT$_{\alpha6}$ \tabularnewline 
			
			\hline
			& $\boldsymbol{\alpha}$  &  \textbf{0.0}  & $1 \times 10^{-5}$ & $2 \times 10^{-5}$ & $2 \times 10^{-6}$  &  $4 \times 10^{-6}$  & $-1 \times 10^{-5}$  & $-3 \times 10^{-5}$  \tabularnewline
			
			Parameters & $\beta$  &  0.0  &  $5 \times 10^{-4}$ &  $5 \times 10^{-4}$ & $5 \times 10^{-4}$  & $5 \times 10^{-4}$  & $5 \times 10^{-4}$ & $5 \times 10^{-4}$  \tabularnewline
			
			\hline
			& $M_{max}$  &  2.183 $M_\odot$ &  2.236 $M_\odot$  & 2.234 $M_\odot$  &  2.237 $M_\odot$  & 2.237 $M_\odot$  & 2.239 $M_\odot$ & 2.243 $M_\odot$\tabularnewline
			
			TM1 %& $R_{M_{max}}$ &  11.22 km  \ \ &  11.64 km   \ \ & 11.70 km \ \ & 11.90 km  \ \ & 12.44 km  \ \ & 11.52 km \ \ & 10.87 km  \tabularnewline
			
			&$R_{1.4}$& 14.322 km  \ \ &  15.221 km \ \ & 14.949 km  \ \ & 15.493 km  \ \ & 15.423 km \ \ & 16.130 km \ \ & 17.234 km  \tabularnewline
			
			&$\Bar{\rho}_{c}$& 5.347 $fm^{-4}$ \ \ & 4.979 $fm^{-4}$ \ \ & 4.982 $fm^{-4}$ \ \ & 4.977 $fm^{-4}$ \ \ & 4.978 $fm^{-4}$ \ \ & 4.971 $fm^{-4}$ \ \ & 4.940 $fm^{-4}$ \tabularnewline
			
			&$C_{max}$& 0.176 \ \ &  0.173  \ \ & 0.174 \ \ & 0.171  \ \ & 0.172  \ \ & 0.169  \ \ & 0.166  \tabularnewline
			
			&$C_{1.4}$& 0.097 \ \ & 0.091  \ \ & 0.093  \ \ & 0.090  \ \ & 0.090  \ \ & 0.086  \ \ & 0.081  \tabularnewline
			\hline
			&  &   \ \ &    \ \ &\ \ &   \ \ &    \ \  &   \ \ &  \ \ \tabularnewline
			
			&  &  General  \ \ &    \ \ &\ \ &   \ \ & Generalized    \ \  &   \ \ &  \ \ \tabularnewline
			&  &  Relativity  \ \ &    \ \ &\ \ &   \ \ & Rastall's gravity \ \  &   \ \ &  \ \ \tabularnewline
			&  &    \ \ &     \ \ &\ \ &   \ \ &    \ \  &   \ \ &  \ \ \tabularnewline
			
			\hline
			&  & \ \ TOV \ \ & GRT$_{\beta1}$ \ \ & GRT$_{\beta2}$ \ \ & GRT$_{\beta3}$ \ \ & GRT$_{\beta4}$ \ \ & GRT$_{\beta5}$ \ \ & GRT$_{\beta6}$ \tabularnewline 
			\hline
			& $\boldsymbol{\beta}$  &  \textbf{0.0}  &  $1 \times 10^{-5}$ &  $1 \times 10^{-4}$  &  $3 \times 10^{-4}$  &  $5 \times 10^{-4}$  &  $7 \times 10^{-4}$  &  $-1 \times 10^{-4}$ \tabularnewline
			
			Parameters & $\alpha$  &  0.0  & $1 \times 10^{-5}$ & $1 \times 10^{-5}$ & $1 \times 10^{-5}$  & $1 \times 10^{-5}$ &  $1 \times 10^{-5}$  & $1 \times 10^{-5}$   \tabularnewline
			\hline
			& $M_{max}$  &  2.183 $M_\odot$ &  2.182 $M_\odot$  & 2.192 $M_\odot$  &  2.213 $M_\odot$  &  2.236 $M_\odot$ & 2.258 $M_\odot$  & 2.171 $M_\odot$\tabularnewline
			
			TM1 %& $R_{M_{max}}$ &  11.22 km  \ \ &  10.09 km   \ \ & 10.87 km \ \ & 12.54 km  \ \ & 13.07 km  \ \  & 13.67 km  \ \ & 9.82 km \ \ \tabularnewline
			
			&$R_{1.4}$& 14.322 km  \ \ &  14.041 km \ \ & 14.250 km  \ \ & 14.730 km  \ \ & 15.222 km  \ \ & 15.749km  \ \ & 13.778 km \ \ \tabularnewline
			
			&$\Bar{\rho}_{c}$& 5.347 $fm^{-4}$ \ \ & 5.351 $fm^{-4}$ \ \ & 5.297 $fm^{-4}$ \ \ & 5.138 $fm^{-4}$ \ \ & 4.979 $fm^{-4}$ \ \ & 4.846 $fm^{-4}$ \ \ & 5.431 $fm^{-4}$ \tabularnewline
			
			&$C_{max}$& 0.176 \ \ &  0.177  \ \ & 0.176 \ \ & 0.174  \ \ & 0.173  \ \ & 0.171  \ \ & 0.178  \tabularnewline
			
			&$C_{1.4}$& 0.097 \ \ & 0.099  \ \ & 0.098  \ \ & 0.095  \ \ & 0.091  \ \ & 0.088  \ \ & 0.101  \tabularnewline
			\hline
			\hline
		\end{tabular}
	}
\end{table*}

%%%%%%%%%%%%%%%%%%%%%%%%%%%%%%%%%%%%%%%%%%%%%%%%%%%%%%%%%%%%%%%%%%%%%
\section*{Acknowledgments}
\noindent 
This work is a part of the project INCT-FNA Proc. No. 464898/2014-5, partially supported by 
Conselho Nacional de Desenvolvimento Cient\'ifico e Tecnol\'ogico (CNPq) under grant No. 301155/2017-8 (D.P.M.). C.E.M. is supported by Coordena\c c\~ao de Aperfei\c coamento de Pessoal de N\'ivel Superior (CAPES) with M.Sc. scholarships. LCNS would like to thank Conselho Nacional de Desenvolvimento Cient\'ifico e Tecnol\'ogico (CNPq) for partial financial support through the research Project No. 164762/2020-5, FMS would like to thank CNPq for financial support through the research Project No. 165604/2020-4 and IPL would like to acknowledge the National Council for Scientific and Technological Development - CNPq grant 306414/2020-1.

%\bibliography{referencias}% Produces the bibliography via BibTeX.

%%%%%%%%%%%%%%%%%%%%%%%%%%%%%%%%%%%%%%%%%%%%%%%%%%%%

\end{document}